\documentclass{aa}
\usepackage{graphicx}

\newcommand{\XMM}{{\it XMM-Newton\ }}
\newcommand{\ROSAT}{{\it ROSAT\ }}
\newcommand{\Chandra}{{\it Chandra\ }}

\newcommand{\wav}{{\tt WAVDETECT}}
\newcommand{\sex}{{\tt SExtractor}}
\newcommand{\ewav}{{\tt EWAVELET}}
\newcommand{\eml}{{\tt EMLDETECT}}
\newcommand{\vtp}{{\tt VTPDETECT}}
\newcommand{\mrsex}{{\tt MR/1+SE}}
\newcommand{\mr}{{\tt MR/1\ }}
\newcommand{\gsex}{{\tt G+SE\ }}

\newcommand{\flux}{erg/s/cm$^2$}

\begin{document}

\thesaurus{}

\title{Comparison of source detection procedures for \XMM images}

\author{I.~Valtchanov \and M.~Pierre \and
R.~Gastaud}
\institute{CEA/DSM/DAPNIA Service d'Astrophysique, 91191
  Gif-sur-Yvette, France}
\date{Received 31 october 2000 / Accepted 8 February 2001}

\offprints{I.~Valtchanov, ivaltchanov@cea.fr}
\titlerunning{Source detection for \XMM}
\maketitle

\begin{abstract}
  
  Procedures based on current methods to detect sources in X-ray
  images are applied to simulated \XMM images. All significant
  instrumental effects are taken into account, and two kinds of sources
  are considered -- unresolved sources represented by the telescope
  PSF and extended ones represented by a $\beta$-profile model.
  Different sets of test cases with controlled and realistic input
  configurations are constructed in order to analyze the influence of
  confusion on the source analysis and also to choose the best methods
  and strategies to resolve the difficulties.  
  
  In the general case of point-like and extended objects the mixed
  approach of multiresolution (wavelet) filtering and subsequent
  detection by \sex\  gives the best results.  In ideal cases of
  isolated sources, flux errors are within 15-20\%. The maximum
  likelihood technique outperforms the others for point-like sources
  when the PSF model used in the fit is the same as in the images.
  However, the number of spurious detections is quite large.
  
  The classification using the half-light radius and \sex\ stellarity
  index is succesful in more than 98\% of the cases. This suggests
  that average luminosity clusters of galaxies ($L_{[2-10]keV} \sim
  3\times 10^{44}$ erg/s) can be detected at redshifts greater than
  1.5 for moderate exposure times in the energy band below 5 keV,
  provided that there is no confusion or blending by nearby sources.
  
  We find also that with the best current available packages,
  confusion and completeness problems start to appear at fluxes around
  $6\times 10^{-16}$ \flux\ in [0.5-2] keV band for \XMM deep surveys.

\end{abstract}

\section{Introduction}

X-ray astronomy has entered a new era now that \Chandra and \XMM are
in orbit.  Their high sensitivities and unprecedented image qualities
bear great promises but also pose new challenges. In this paper, we
outline problems of object detection in X-ray images that were not
previously encountered. In doing so, we compare the performances of
various detection techniques on simulated \XMM test images,
incorporating the main instrumental characteristics.

The X-ray observations consist of counting incoming photons one by
one, recording their time of arrival, position and energy.  Later, the
event list is used to create images for a given pixel scale and energy
band.  Various X-ray telescope effects complicate this simple picture
-- the point spread function (PSF) and the telescope effective area
(the vignetting effect), both dependent on the off-axis angle and
incoming photon energy; detector effects like quantum efficiency
variations, different zones not exposed to X-ray photons;
environmental and background effects like solar flares and particle
background.  Even for relatively large exposures, the X-ray images
could contain very few photons, and some sources could contain only a
few tens of photons spread over a large area.  Consequently, it is
important for the source detection and characterization procedures to
be able to cope with these difficulties.

\begin{figure}
  \centerline{
    \includegraphics[width=7cm]{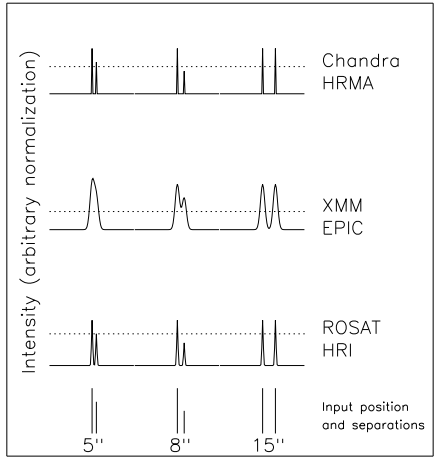}}
  \caption{
    Typical representation of objects seen by \ROSAT-HRI, \XMM-EPIC
    and \Chandra-HRMA. The objects are represented by
    $\delta$-functions and folded by the corresponding instrumental
    PSF and efficiency.  Full-width-at-half-maximum (FWHM) of the
    on-axis PSFs are $1\farcs 7$ for \ROSAT-HRI, $0\farcs 5$ for
    \Chandra-HRMA and $6\arcsec$ for \XMM-EPIC.  The dotted line
    represents schematically the detection limit.}
  \label{fig:sat}
\end{figure}

As an example, the same hypothetical input situation is shown
schematically in Fig.~\ref{fig:sat} for
\ROSAT\footnote{http://wave.xray.mpe.mpg.de/rosat},
\XMM\footnote{http://xmm.vilspa.esa.es/} and
\Chandra\footnote{http://chandra.harvard.edu/} . \XMM's rather large
PSF, coupled with its higher sensitivity, leads to the detection of
more objects but also to blending and source confusion, which become
severe for long exposures depending on the energy band.  Confusion
problems in the hard band above 5 keV are less important, given the
smaller number of objects and smaller count rate of energetic photons.
Thus, why we concentrate our analysis mainly on source detection
problems for the more complicated case of the \XMM energy bands below
5 keV.

Each X-ray mission provides data analysis packages -- EXSAS for \ROSAT
(Zimmermann et al. \cite{exsas}), CIAO for \Chandra (Dobrzycki et al.
\cite{ciao}) and \XMM Science Analysis System
(XMM-SAS\footnote{http://xmm.vilspa.esa.es/sas/}).  They include
procedures for source detection, and in this paper we estimate and
compare their performances on simulated images using various types of
objects. These procedures make use of techniques such as Maximum
Likelihood (ML), Wavelet Transformation (WT), Voronoi Tessellation and
Percolation (VTP).

In Section \ref{sec:sim} we describe the X-ray image simulations. A
short presentation of the detection procedures is given in
Sec.~\ref{sec:det}. Tests using only point sources are presented in
Sec.~\ref{sec:test1}, and extended sources in Sec.~\ref{sec:test2}.
We have analyzed realistic simulations of a shallow and a deep
extragalactic field with only point sources in Sec.~\ref{sec:test3}
and with extended objects in Sec.~\ref{sec:test4} for an exposure of
10 ks. Finally, we investigate the problems of confusion and
completeness in two energy bands -- [0.5-2] and [2-10] keV for two
exposures -- 10 ks and 100 ks (Sec.~\ref{sec:test5}).
Sec.~\ref{sec:disc} presents the conclusions. (H$_0=50$ km/s/Mpc,
$h=0.5$, q$_0=0.5$ and $\Omega_0=1$ are used).

\section{Simulation of X-ray images}
\label{sec:sim}

The simulations are essential to understand and qualify the behavior
of the different detection and analysis packages.  We have developed a
simulation program that generates X-ray images for given exposure
times with extended and point-like objects. It takes into account the
main instrumental characteristics of \XMM and the total sensitivity of
the three EPIC instruments.  The procedure is fast and flexible and is
made of two independent simulation tasks: object generation
(positions, fluxes, properties) and instrumental effects.  A
possibility to apply the instrumental response directly over images is
also implemented, especially useful when one wants to use sky
predictions from numerical simulations (cf. Pierre et al.
\cite{mp00}).

A summary of the simulated images parameters are given in
Tab.~\ref{tab:sim}.

\begin{table}
\caption{The general parameters for the simulated images.}
\label{tab:sim}
\begin{tabular}{ll}
\hline\hline
Parameter \\
\hline
Image scale                             & $4\arcsec$/pixel \\
Image size                              & 512x512 \\
Exposure time                           & 10ks \& 100ks\\
Energy bands                            & [0.5-2]  \& [2-10] keV\\
PSF on axis                             & $6\arcsec$ (FWHM)\\
                                        & $15\arcsec$ (HEW)\\
\multicolumn{2}{c}{Total background (pn+2MOS)} \\
$[0.5-2]$ keV                             & $1.78 \times 10^{-5}$ cts/s/pixel \\
                                        & 0.0041 cts/s/arcmin$^2$ \\
$[2-10]$ keV                              & $2.4 \times 10^{-5}$ cts/s/pixel \\
                                        & 0.0055 cts/s/arcmin$^2$ \\
\hline
\end{tabular}
\end{table}

The point-like sources are assumed to be AGNs or QSOs with a power law
spectrum with a photon index of 2 and flux distribution following the
$\log N-\log S$ relations of Hasinger et al. (\cite{has98},
\cite{has01}) and Giacconi et al. (\cite{gia01}) in the two energy
bands.

The PSF model is derived from the current available calibration
files\footnote{http://xmm.vilspa.esa.es/ccf/}.  On-board PSF data is
generally in very good agreement with the previous ground based
calibrations (Aschenbach et al.  \cite{asch00}).  We must stress that
the model PSF is an azimuthal average and in reality, especially for
large off-axis angles, its shape can be very distorted. However, in
the analytic model (Erd et al.  \cite{calbook}), the off-axis and
energy dependences are not available yet. This is not crucial, as the
energy dependence in the bands used is moderate and we confine all
the analysis inside $10\arcmin$ from the centre of the field-of-view
where the PSF blurring is negligible.

The extended objects are modeled by a $\beta$-profile (Cavaliere \&
Fusco-Femiano \cite{cff76}) with fixed core radius $r_c=250\,h^{-1}$
kpc and $\beta=0.75$. A thermal plasma spectrum (Raymond \& Smith
\cite{rs77}) is assumed for different temperatures, luminosities and
redshifts.

The source spectra (extended and point-like) are folded with the
spectral response function for the total sensitivity of the three \XMM
EPIC instruments (MOS1, MOS2 and pn with thin filters) by means of
XSPEC (Arnaud \cite{xspec}) to produce the count rates in different
energy bands.  The actual choice of the energy bands is not important
for this comparison study, although some objects can be more
efficiently detected in particular energy ranges.

As an example, we show in Tab.~\ref{tab:counts} the resulting count
rates for extended sources assuming that they represent an average
cluster of galaxies.

\begin{table}
  \caption{
    Count rates for extended sources in three energy bands calculated
    assuming an average luminosity ($L_{[2-10]keV}=2.8\times 10^{44}$
    erg/s, $T_X=5$ keV) $\beta-$profile cluster of galaxies with
    a Raymond-Smith spectrum, $N_H=5\times 10^{20}$ cm$^{-2}$ and spectral
    response functions for the three EPIC detectors with thin filters.  }
  \label{tab:counts}
  \begin{tabular}{cclll}
    \hline\hline
    \multicolumn{1}{c}{z} & 
    \multicolumn{1}{c}{Core radius} &
    \multicolumn{3}{c}{Count-rate, photons/s } \\
    &  
    \multicolumn{1}{c}{arcsec} &
    \multicolumn{1}{c}{[0.4-4] keV} & 
    \multicolumn{1}{c}{[0.5-2] keV} & 
    \multicolumn{1}{c}{[2-10] keV} \\
    \hline
    0.6      &  32.8 & 0.1687 &  0.1316 &  0.0362        \\
    0.7      &  31.3 & 0.1238 &  0.0963 &  0.0253        \\
    0.8      &  30.4 & 0.0942 &  0.0734 &  0.0185        \\
    0.9      &  29.7 & 0.0737 &  0.0577 &  0.0139        \\
    1.0      &  29.3 & 0.0593 &  0.0465 &  0.0107       \\
    1.1      &  29.1 & 0.0486 &  0.0382 &  0.0085       \\
    1.2      &  29.0 & 0.0406 &  0.0319 &  0.0068       \\
    1.3      &  29.0 & 0.0343 &  0.0270 &  0.0055       \\
    1.4      &  29.1 & 0.0293 &  0.0231 &  0.0046       \\
    1.5      &  29.2 & 0.0253 &  0.0200 &  0.0038       \\
    1.6      &  29.4 & 0.0220 &  0.0175 &  0.0032       \\
    1.7      &  29.6 & 0.0193 &  0.0154 &  0.0027       \\
    1.8      &  29.9 & 0.0171 &  0.0137 &  0.0023       \\
    1.9      &  30.2 & 0.0152 &  0.0122 &  0.0020       \\
    2.0      &  30.5 & 0.0137 &  0.0109 &  0.0018       \\
    \hline            
  \end{tabular}
\end{table}

The background in the simulations includes realistic cosmic diffuse
and background values derived from the \XMM in-orbit measurements in
the Lockman Hole (Watson et al. \cite{wat00}).

The calculated count rates for the objects and the photons of the
background are subject to the vignetting effect -- some photons are
lost due to the smaller telescope effective area at given off-axis
angle, depending on the incoming photon's energy. We have parametrized
the vignetting factor -- the probability that a photon at an off-axis
angle $\theta$ to be observed -- as polynomials of fourth order in two
energy bands: [0.5-2] and [2-10] keV, using the latest \XMM on-flight
calibration data. For example, a photon at $\theta=10\arcmin$ has a
53\% chance of being observed in [0.5-2] keV and 48\% in [2-10] keV.

Further instrumental effects such as quantum efficiency difference
between the CCD chips, the gaps between the chips, out-of-time events,
variable background, pile-up of the bright sources are not taken into
account -- their inclusion is not relevant for our main objective.

\section{Detection procedures}
\label{sec:det}

Without attempting to provide a review of the available techniques in
the literature, we briefly describe here the procedures we have
tested.  They are summarized in Tab.~\ref{tab:proc}.

\begin{table*}
  \caption[]{Designation, implementation and short description of the
    procedures or method used for detection and analysis.}
  \label{tab:proc}
  \begin{tabular}{lccp{6cm}}
    \hline
    \hline
    Procedure & Implementation & Version & Method \\
    \hline
    \eml\     & XMM-SAS v5.0       & 3.7.2 & Cell detection + Maximum
    likelihood \\
    \vtp\     & \Chandra CIAO   & 2.0.2   & Voronoi Tessellation and
    percolation \\
    \wav    & \Chandra CIAO   & 2.0.2   & Wavelet\\
    \ewav   & XMM-SAS v5.0       & 2.4 & Wavelet\\
    \gsex   & Gauss + \sex    & 2.1.6 & Mixed -- gauss convolution
    followed by \sex\  detection\\
    \mrsex  & MR/1 + \sex    & 2.1.6 & Mixed -- multi-resolution filtering
    followed by \sex\  detection\\
    \hline
  \end{tabular}
\end{table*}

\subsection{Sliding cell detection and maximum likelihood (ML)
  method}

Historically, the sliding cell detection method was first used for
{\it Einstein Observatory} observations (e.g. EMSS -- Gioia et al.
\cite{emss}). It is included in \ROSAT, \Chandra and \XMM data
analysis tools and a good description can be found in the specific
documentation for each of those missions.

The X-ray image is scanned by a sliding square box and if the
signal-to-noise of the source centered in the box is greater than the
specified threshold value it is marked as an object. The signal is
derived from the pixel values inside the cell and noise is estimated
from the neighboring pixels. Secondly, the objects and some zone
around them are removed from the image forming the so-called
``cheese'' image which is interpolated later by a suitable function
(generally a spline) to create a smooth background image.  The
original image is scanned again but this time using a threshold from
the estimated background inside the running cell to give the map
detection object list.

The procedure is fast and robust and does not rely on {\it a priori}
assumptions. However it has difficulties, especially in detecting
extended features, close objects and sources near the detection limit.
Many refinements are now implemented improving the sliding cell
method: (1) consecutive runs with increasing cell size, (2) matched
filter detection cell where the cell size depends on the off-axis
angle.  However, the most important improvement was the addition of
the maximum likelihood (ML) technique to further analyze the detected
sources.

The ML technique was first applied to analyze \ROSAT observations
(Cruddace et al.  \cite{ml88}, \cite{cru91}, Hasinger et al.
\cite{has93}). It was used to produce all general X-ray surveys from
\ROSAT mission (e.g.  RASS -- Voges et al.  \cite{rass}, WARPS survey
-- Ebeling et al. \cite{warps}).  The two lists from local and map
detection passes can be merged to form the input objects list for the
ML pass. It is useful to feed the ML procedure with as many candidate
objects as possible, having in mind that large numbers of objects
could be very CPU-expensive.  The spatial distribution of an input
source is compared to the PSF model -- the likelihood that both
distributions are the same -- is calculated with varying the input
source parameters (position, extent, counts) and the corresponding
confidence limits can be naturally computed. A multi-PSF fit is also
implemented which helps in deblending and reconstructing the
parameters of close sources. In the output list, only sources with a
likelihood above a threshold are kept.

The ML method performs well and has many valuable features, however,
it has some drawbacks -- it needs a PSF model to perform the
likelihood calculation and thus favours point-like source analysis,
extent likelihood could be reliably taken only for bright sources, it
cannot detect objects which are not already presented in the input
list (e.g.  missing detections in the local or map passes).

Here we have used \eml\ -- an implementation of the method
specifically adapted for XMM-SAS (Brunner \cite{b96}). In the map mode
sliding cell pass we used a low signal-to-noise ratio ($\sim 3\sigma$)
above the background in order to have as many as possible input
objects for the ML pass. The likelihood limit (given by $L = -\ln P$,
where $P$ is the probability of finding an excess above the
background) was taken to be 10, which corresponds roughly to $4\sigma$
detection. A multi-PSF fitting mode with the maximum of 6 simultaneous
PSF profile fits was used.

\subsection{VTP}

VTP -- the Voronoi Tessellation and Percolation method (Ebeling \&
Wiedenmann \cite{vtp1}, Ebeling \cite{vtp2}) is a general method for
detecting structures in a distribution of points (photons in our case)
by choosing regions with enhanced surface density with respect to an
underlying distribution (Poissonian in X-ray images). It treats the
raw photon distribution directly without any recourse to a PSF model
or a geometrical shape of the objects it finds. Each photon defines a
centre of a polygon in the Voronoi tessellation image and the surface
brightness is simply the inverse area of the polygon (assuming one
single photon per cell). The distribution function of the inverse
areas of all photons is compared to that expected from a Poisson
distribution and all the cells above a given threshold are flagged and
percolated, i.e. connected to form an object. This method was
successfully used with \ROSAT data (Scharf et al. \cite{s97}) and is
currently incorporated in the \Chandra DETECT package (Dobrzycki et
al.  \cite{ciao}).

Apart from these advantages, VTP has some drawbacks which are
especially important for \XMM observations: (1) because of the
telescope's high sensitivity and rather large PSF with strong tails,
the percolation procedure tends to link nearby objects; (2) excessive
CPU time for images with relatively large number of photons; (3) there
is no simple way to estimate the extension of objects.

\subsection{Wavelet detection procedures}

In the past few years a new approach has been extensively used: the
wavelet technique (WT). This method consists in convolving an image
with a wavelet function:
\begin{equation}
w_a (x,y) = I(x,y) \otimes \psi(\frac{x}{a},\frac{y}{a}),
\end{equation}
where $w_a$ are the wavelet coefficients corresponding to a scale $a$,
$I(x,y)$ is the input image and $\psi$ is the wavelet function.  The
wavelet function must have zero normalization and satisfy a simple
scaling relation
\begin{equation}
\psi_a(x,y) = \frac{1}{a}\psi_1(\frac{x}{a},\frac{y}{a}).
\end{equation}
The choice of $\psi$ is dictated by the nature of the problem but most
often the second derivative of the Gaussian function or the so called
``Mexican hat'' is used.

The WT procedure consists of decomposing the original image into a
given number of wavelet coefficient images, $w_a$, within the chosen
set of scales $a$. In each wavelet image, features with characteristic
sizes close to the scale are magnified and the problem is to mark the
significant ones, i.e. those which are not due to noise.  In most
cases, this selection of significant wavelet coefficients cannot be
performed analytically because of the redundancy of the WT introducing
cross-correlation between pixels.  For Gaussian white noise, $w_a$ are
distributed normally, allowing easy thresholding. This is not the case
for X-ray images which are in the Poissonian photon noise regime.

Various techniques were developed for selecting the significant
wavelet coefficients in X-ray images. In Vikhlinin et al.
(\cite{vik97}) a local Gaussian noise was assumed; Slezak et al.
(\cite{sle94}) used the Ascombe transformation to transform an image
with Poissonian noise into an image with Gaussian noise; in Slezak et
al.  (\cite{sle93}), Starck \& Pierre (\cite{sp98}) a histogram of the
wavelet function is used.  In recent years a technique based on Monte
Carlo simulations is used successfully (e.g.  Grebenev et al.
\cite{gre95}, Damiani et al. \cite{dam97}, Lazzati et al. \cite{l99}).

Once the significant coefficients at each scale are chosen, the local
maxima at all scales are collected and cross-identified to define
objects.  Different characteristics, such as centroids, light
distribution etc., can be computed, as well as an indication of the
source size at the scale where the object wavelet coefficient is
maximal.

WT has many advantages -- the multiresolution approach is well suited
both for point-like and extended sources but favours circularly
symmetric ones. Because of the properties of the wavelet function a
smoothly varying background is automatically removed. Extensive
description of wavelet transform and its different applications can be
found in Starck et al. (\cite{sta98}).

In this work we have tested two WT procedures:
\begin{description}
\item[\wav:] one of the \Chandra wavelet-based detection techniques
  (Freeman et al.  \cite{fre96}, Dobrzycki et al.  \cite{ciao}). It
  uses a ``Mexican hat'' as wavelet function and identifies the
  significant coefficients by Monte Carlo simulations.  The background
  level needed to empirically estimate the significance is taken
  directly from the negative annulus of the wavelet function.  Source
  properties are computed inside the detection cell defined by
  minimizing the function $|\log r_{PSF} - \log \sigma_{F}|$, where
  $r_{PSF}$ is the size of the PSF encircling a given fraction of the
  total PSF flux and $\sigma_{F}$ is the size of the object at the
  scale closest to the PSF size.  No detailed PSF shape information is
  needed to perform this minimization -- just $r_{PSF}$ as a function
  of the off-axis angle. The control parameters which need attention
  are the significance threshold and the set of scales. We have used a
  significance threshold of $10^{-4}$ corresponding to 1 false event
  in 10000 ($\sim 4\sigma$ in Gaussian case) and ``$\sqrt{2}$
  sequence'' for the scales, where
  $a=1,\sqrt{2},2,2\sqrt{2},4,3\sqrt{2}\dots 16$.
\item[\ewav:] XMM-SAS package, based on a wavelet analysis. It is very
  similar to \wav\ but implements some new ideas. The sources are
  assumed to have a Gaussian shape in order to analytically derive
  their counts and extent. Currently, the PSF information is ignored,
  which can be regarded as a serious drawback. We have used the
  significance threshold of $10^{-4}$ ($\sim4\sigma$) and wavelet
  scales 1, 2, 4, 8 and 16.
\end{description}

\subsection{Mixed approach}

This method combines a source detection by an elaborated procedure
over filtered/smoothed raw photon images.  

The use of such a mixed approach is motivated by the fact that
procedures for source detection in astronomical images have been
developed for many years and the steps and problems of deblending,
photometry, classification of objects are now quite well understood.
The raw photon image manipulations can be performed with very simple
smoothing procedures (for example a Gaussian convolution) or with more
sophisticated methods like the ``matching filter'' technique, adaptive
smoothing or multiresolution (wavelet) filtering.

We have used two different types of raw image filtering:

\begin{description}
\item[(1)] Gaussian convolution. For our simulated images we applied
  two convolutions with FWHM=$12\arcsec$ and $20\arcsec$ constrained
  by the image characteristics (see Sec.~\ref{sec:sim}).
\item[(2)] Multiresolution iterative threshold filtering (\mr\ 
  package, Starck et al. \cite{sta98}) by means of ``\`a trous'' (with
  holes) wavelet method and Poissonian noise model (Slezak et al.
  \cite{sle93}, Starck \& Pierre \cite{sp98}) to flag the significant
  wavelet coefficients (also known as auto-convolution or wavelet
  function histogram method).  The control parameters are the
  significance threshold, wavelet scales and the number of iterations
  for the image reconstruction. We took $10^{-4}$ significance
  threshold (corresponding to $\sim 4\sigma$ for Gaussian
  distribution), 6 scales for the ``\`a trous'' method allowing
  analysis of structures with characteristic sizes up to $2^6$ pixels
  ($\sim 4\arcmin$). We have used 25 iterations or a reconstruction
  error less than $10^{-5}$.  More iterations can improve the
  photometry but can also lead to the appearance of very faint real or
  sometimes false features.  Optionally the last-scale wavelet
  smoothed image in the iterative restoration process is used for
  analysis of large-scale background variations or very extended
  features.
\end{description}

Source detection over the filtered images was performed with \sex\ 
(Bertin \& Arnouts \cite{sex}) -- one of the most widely used
procedures for object detection in astronomy with a simple interface
and very fast execution time. Originally, \sex\ was developed for
optical images but it can be applied very successfully on X-ray images
after they are properly filtered. The flexibility of \sex\ is ensured
by a large number of control parameters requiring dedicated
adjustments. In the following we will briefly outline those having a
significant influence on our results:

\begin{itemize}
\item Detection and analysis threshold -- important parameters for
  Gaussian convolved images. It is preferable to use a value greater
  than $4\sigma$, even sacrificing very faint sources but reducing the
  number of false detections. For \mr\ filtering this is irrelevant
  because the features in the images are already at $4\sigma$
  significance.
\item Background mesh size -- influences object detection as well as
  photometry. Using too coarse a background grid (less grid points)
  will lead to a smoother background image but some faint objects will
  be missed in the detection. While too detailed a map (more grid
  points) will lead to many false detections and worse photometry,
  depending on the local noise properties. We have found after many
  experiments that local background meshes with 32 to 64 nodes give
  best results balancing both effects.
\item Minimum detection area -- in some cases this parameter can help
  to avoid spurious detections especially when one uses wavelet scales
  below the characteristic size of objects.
\item Deblending parameters -- the number of the deblending levels and
  the minimum contrast. We used the extreme values of 64 levels and
  zero contrast so that any saddle point will lead to object splitting.
\item Photometry -- we have decided to use the automatic aperture
  photometry procedure implemented in \sex\ following Kron
  (\cite{kro80}) and Infante (\cite{inf87}) which gives better results
  than the fixed aperture or isophotal methods.
\item Classification -- we discuss the classification in the next
  subsection.
\end{itemize}

\subsection{Classification}
\label{sec:class}

The problem of classifying sources as resolved or unresolved (or
stars/galaxies, extended/point-like) has a long history and
discussions can be found in widespread detection packages like {\tt
  FOCAS} (Valdes et al. \cite{focas}), {\tt INVENTORY} (Kruszewski
\cite{inventory}), \sex\ (Bertin \& Arnouts \cite{sex}), Neural
Extractor (Andreon et al.  \cite{next}). This task becomes even more
difficult for X-ray images -- we have already mentioned various X-ray
telescope effects that blur and distort the images as a function of
the off-axis angle.
\begin{description}
\item[\eml]-- the extension likelihood is calculated by the deviation
  of the object fitted from the PSF model when varying the size
  (extension) of the object. It was used with \ROSAT observations to
  flag possible extended objects and in most of the cases the results
  were positive.  The extension likelihood depends on the chosen PSF
  model and for faint objects is not reliable (De Grandi et al.
  \cite{sdg97}).
\item[\wav]-- the ratio between object size and the PSF size
  ($R_{PSF}$) is used for classification purposes.  It depends on the
  adopted definition for object size and can be misleading, especially
  for faint objects.
\item[\sex] -- one parameter for classification is the stellarity
  index. It is based on neural-network training performed over set of
  some $10^6$ simulated optical images. There are 10 parameters
  included in the neural network training -- 8 isophotal areas, the
  maximal intensity and the ``seeing''. The only controllable
  parameter is the ``seeing'' which, however, has no meaning for space
  observations.
  It can be tuned to the mean PSF size.\\
  The half-light radius $R_{50}$ is another classification parameter.
  It can be used as a robust discriminator because it depends only on
  the photon distribution inside the object (``luminosity profile''),
  which is assumed to be different for resolved and unresolved
  objects.
\end{description}
Additional indications might be taken into account when the
classification is ambiguous -- e.g. the wavelet scale at which
object's wavelet coefficient is maximal or even spectral signatures.


\section{Test~1 -- point-like sources}
\label{sec:test1}

\subsection{Input configuration} 

We address the problem of point-like sources separated by $15\arcsec$
(half-energy width of the on-axis PSF), $30\arcsec$ and 60$\arcsec$
with different flux ratios. We include the PSF model and background
but do not apply the vignetting effect.

The raw input test image is shown in Fig.~\ref{test1:sum} together
with its Gaussian convolution, \mr\  wavelet filtering and \wav\ output
image.  Visually, the Gaussian image is quite noisy, while there are
few spurious detections in the WT images.

\begin{figure*}
\vspace{14cm}
  \caption{
    Test~1. The raw X-ray photon image with object counts for 10 ks
    exposure time (upper left). There are three sets of two columns of
    objects at separations of $15\arcsec$ (HEW of the PSF),
    $30\arcsec$ and $60\arcsec$. The Gaussian convolution with
    FWHM$=12\arcsec$ (upper right), \mr\ wavelet filtering (lower
    left) and \wav\ detection image (lower right) both with a
    significance threshold of $10^{-4}$ are shown.}
  \label{test1:sum}
\end{figure*}

\subsection{Detection rate and positional errors}

The number of missing detections and false objects are shown in
Tab.~\ref{test1:tab}.

\begin{table}
  \caption{Test~1. Detection results. The total number of input
    objects is 36.}
  \label{test1:tab}
  \begin{tabular}{rcc}
    \hline\hline
    Method & Missed & False \\
    \hline
    \eml & 4 & 13\\
    \gsex ($4\sigma$) & 6 & 1 \\
    \mrsex & 7 & 1 \\
    \wav   & 7 & 21 \\
    \ewav & 6 & 4 \\
    \vtp\  & 12 & 19 \\
    \hline
  \end{tabular}
\end{table}

The one sigma input-detect position differences are less than the FWHM
of the PSF ($6\arcsec$) for all procedures and the maximum occurs for
the blended objects, as expected. Note the large number of spurious
detections with \wav, \vtp\  and \eml.

\subsection{Photometric accuracy} 

\begin{figure*}
  \centerline{\hbox{
      \includegraphics[width=7cm]{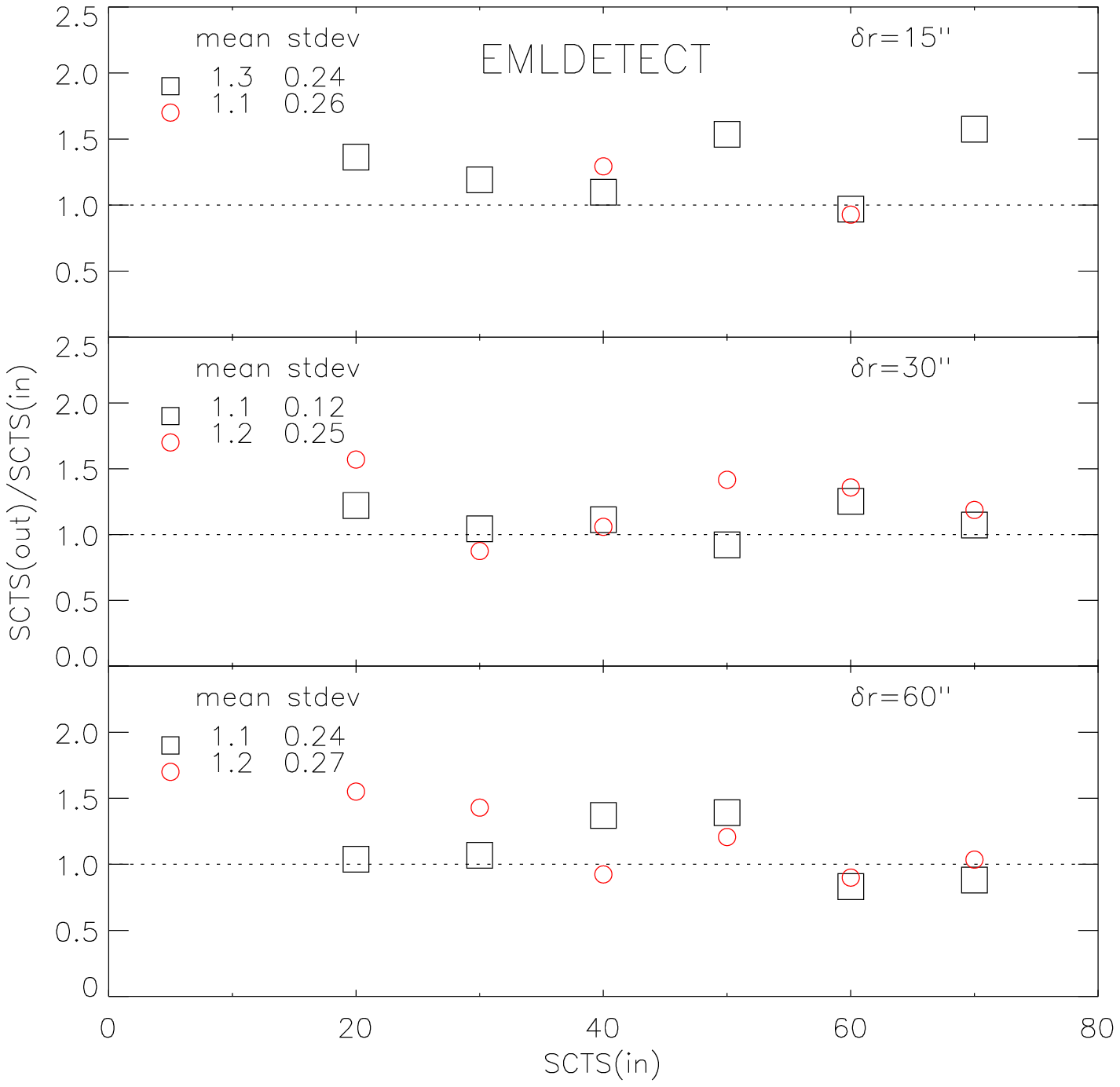} \hfil
      \includegraphics[width=7cm]{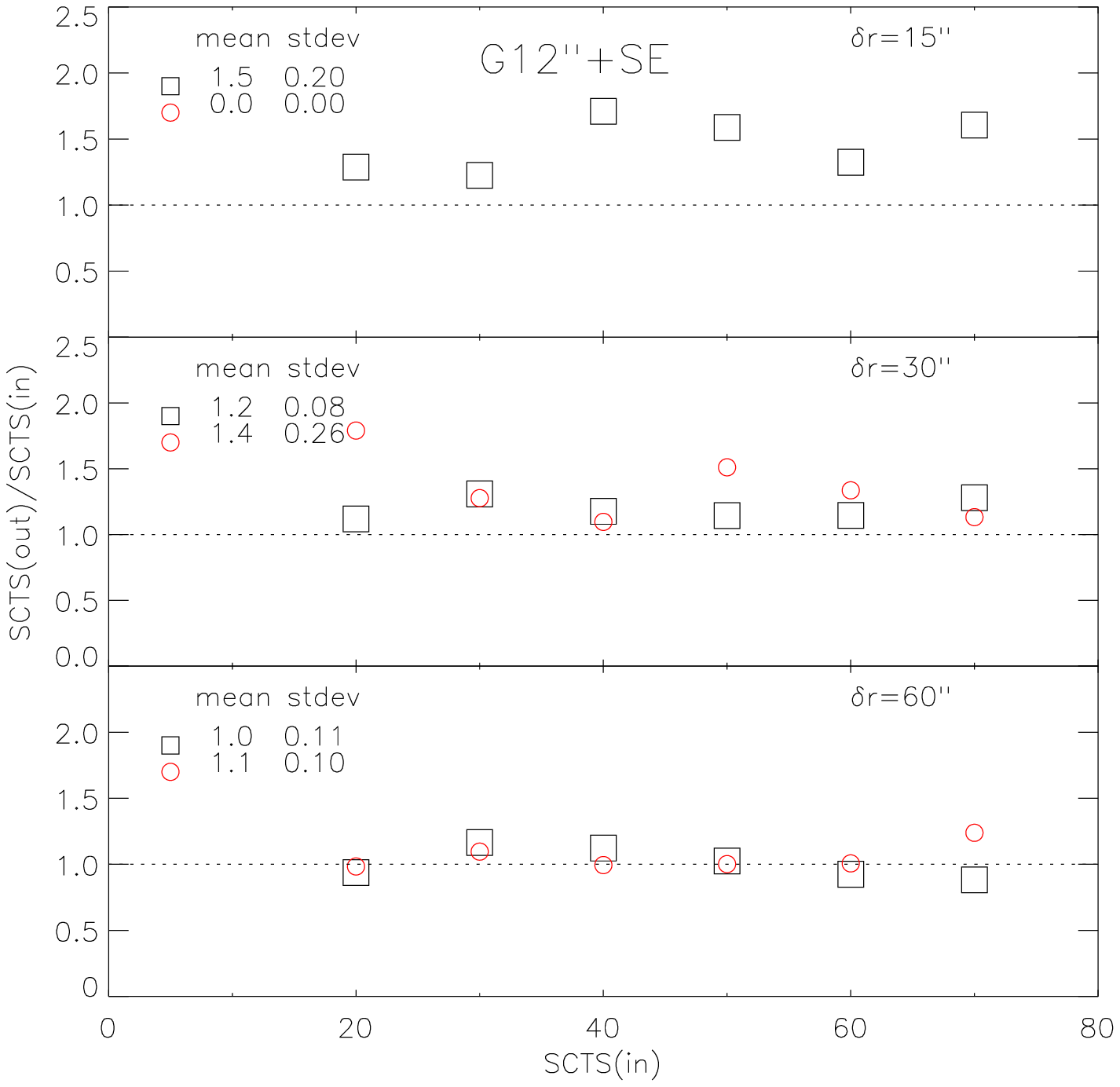}
      }}
  \centerline{\hbox{
      \includegraphics[width=7cm]{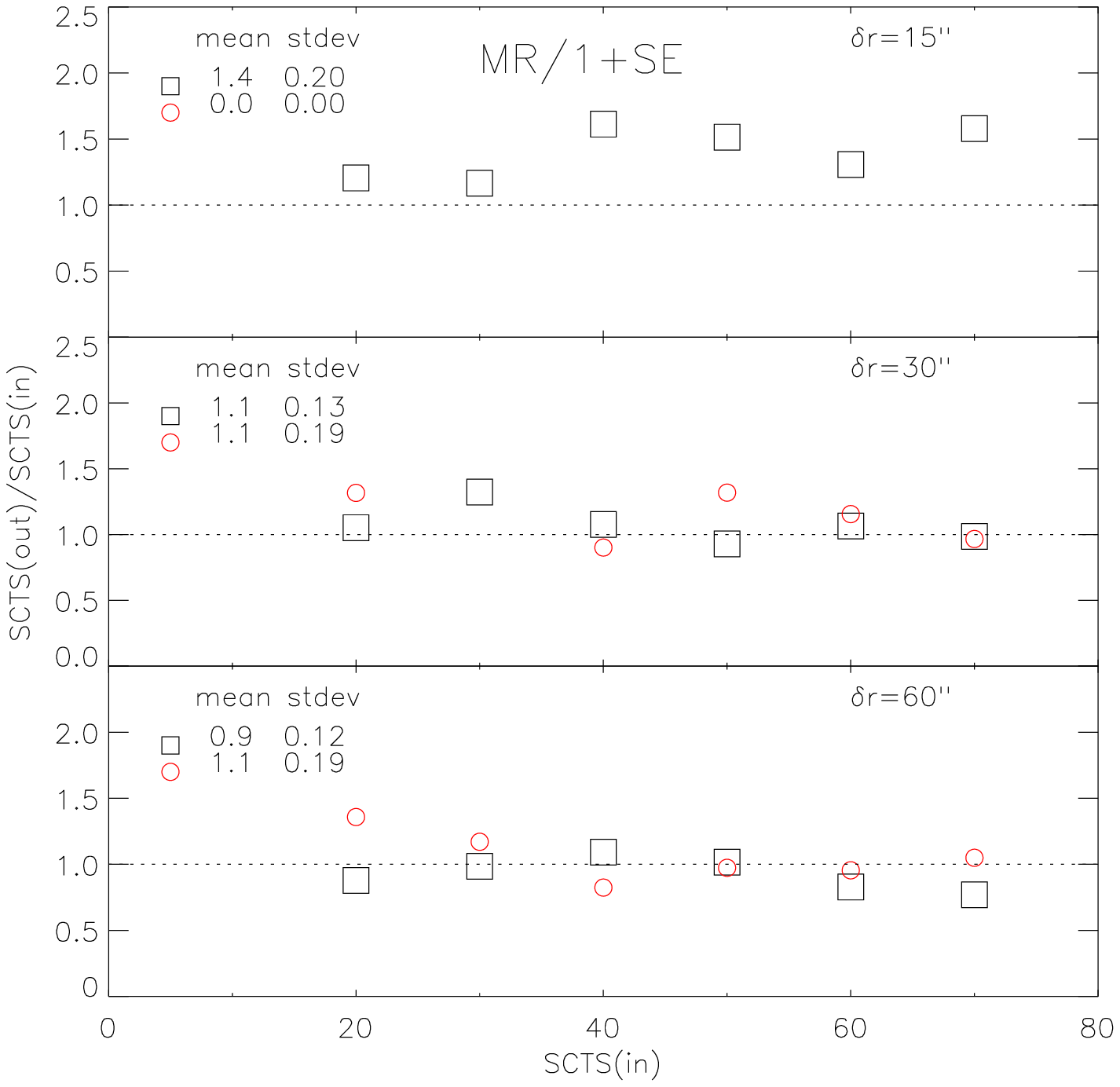} \hfil
      \includegraphics[width=7cm]{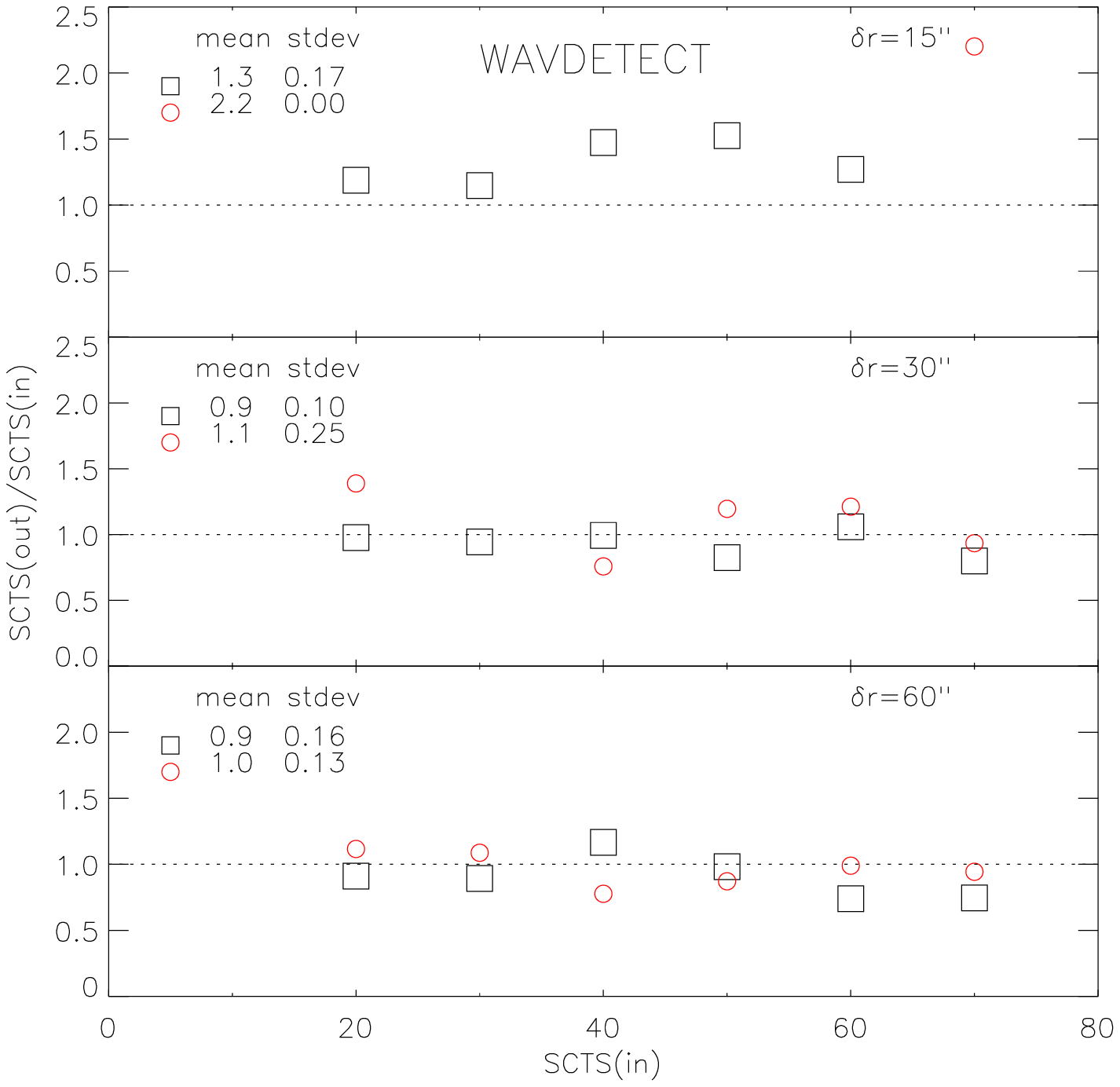}
      }}
  \centerline{\hbox{
      \includegraphics[width=7cm]{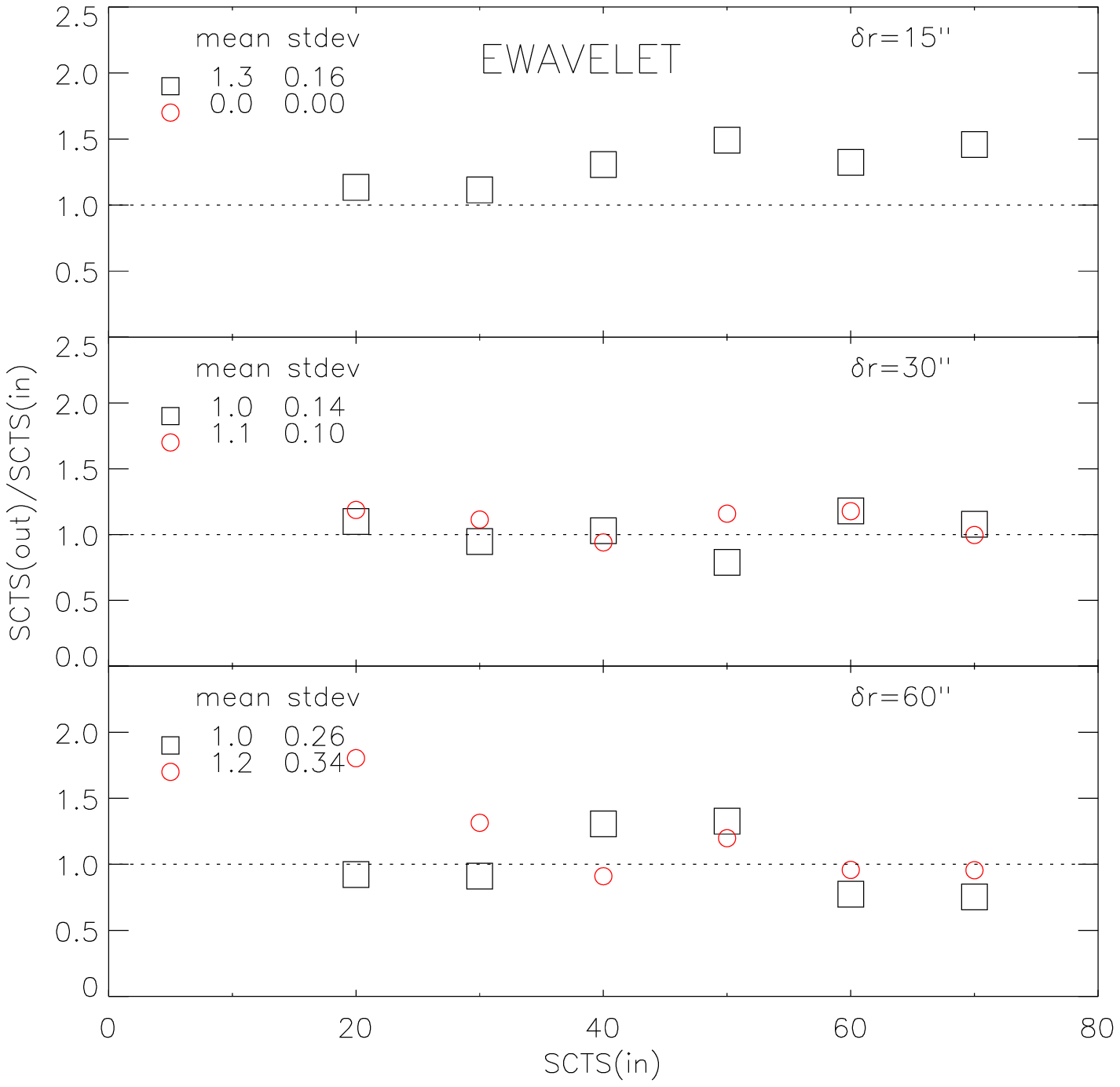} \hfil
      \includegraphics[width=7cm]{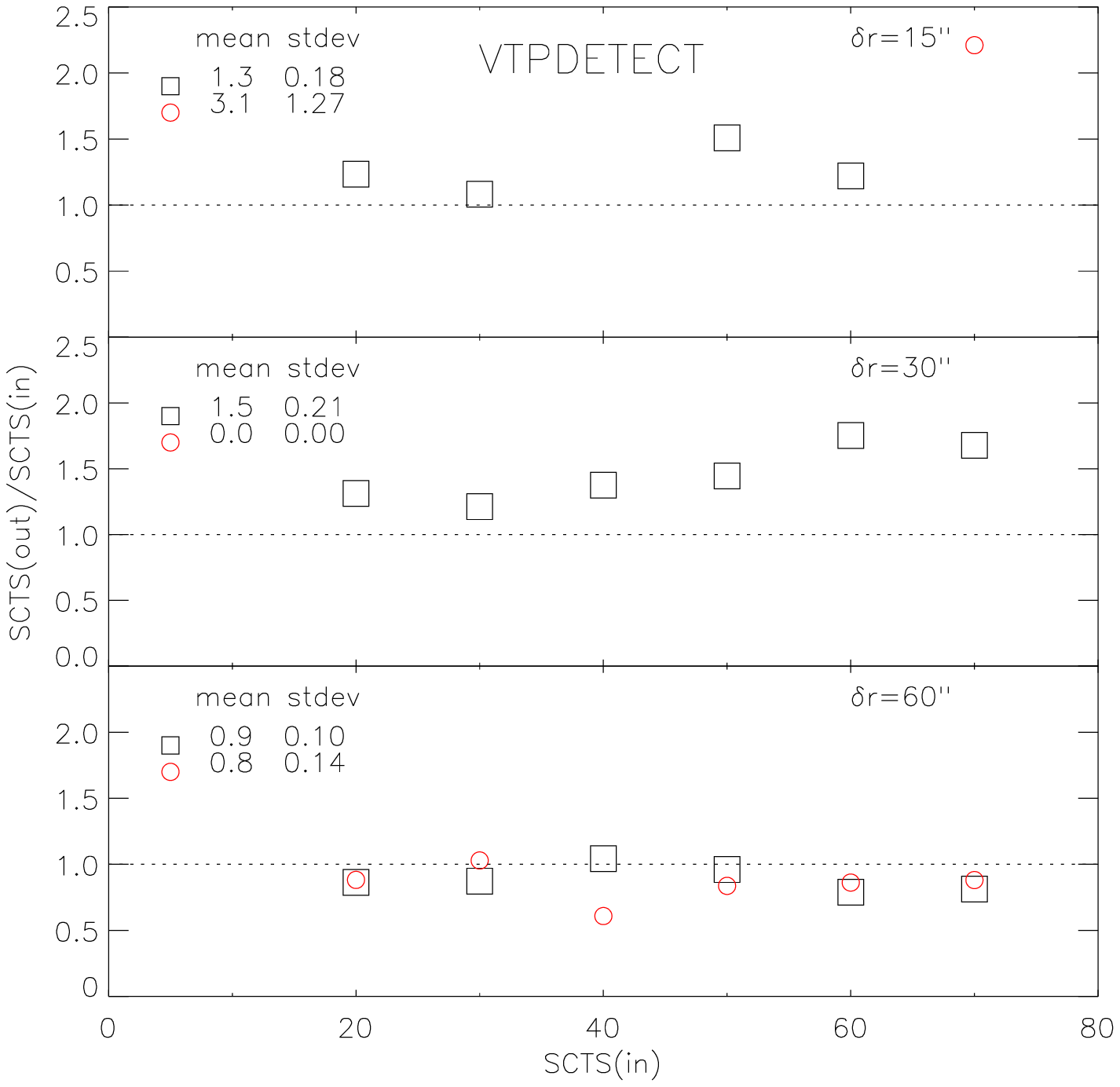}
      }}
  \caption{
    Test~1. The three panels of each figure show the results in terms
    of the ratio of inferred counts $SCTS(out)$ and input counts
    $SCTS(in)$ as a function of the varying input counts for the three
    cases of object separations (indicated by $\delta r$). The mean
    and st.dev.  of the corresponding points are also indicated.  When
    there are no detections, the mean and the st.dev. are both zero.
    Objects with input counts fixed at 100 (squares) are placed beside
    their corresponding neighbors (circles), rather than being plotted
    at 100.}
  \label{test1:results}
\end{figure*}

The results for the photometry in terms of the inferred to the input
counts are shown in Fig.~\ref{test1:results}.

\begin{description}
  
\item[$\delta r=15\arcsec$.] Only \eml\ detects two of the six fainter
  objects. None of the other procedures separates the objects and
  consequently the inferred counts are a blend from both sources.
  
\item[$\delta r=30\arcsec$.] The proximity of objects influences the
  detection and the photometry. \eml\ and \ewav\ show the best
  detection rate results while all the other procedures miss one of
  the faintest objects.
  
\item[$\delta r=60\arcsec$.] We can safely assume that the objects are
  well separated. The recovery of the properties is informative for
  the performance of the tested procedures. It is clear that the
  general flux reconstruction error (taken to be the spread of the
  points around the unity line in Fig.~\ref{test1:results}) is about
  15\% for brighter sources and goes down to 20-25\% for the faintest
  ones.  [In our 10 ks exposure tests and with the adopted background
  in band [0.5-2] keV, we assume the objects with input counts of 20
  photons ($\sim 10^{-15}$ \flux) to be at the detection limit when
  there is no confusion by nearby sources.]
\end{description}

\subsection{Discussion} 

After this simple test we can eliminate the \vtp: in addition to the
very large execution time, some of the VTP-detected object centres
were shifted by more than $20\arcsec$ from their input positions -- a
consequence of its ability to detect sources with different shapes
where the object center can be far from the input position.  Moreover,
\vtp\ percolates all the double sources into single objects at $\delta
r=30\arcsec$, which all other procedures were able to separate.

No procedure unambiguously shows best results -- both in terms of the
detection rate, spurious sources and photometric reconstruction. \eml\ 
outperforms the others in terms of detection rate but with the price
of many spurious detections. Using exactly the same PSF model as the
one hard-coded in \eml\ leads to much better photometric
reconstruction.

All other procedures are comparable: \ewav\ showing better detection
but its photometric reconstruction is far from satisfactory -- about
half of the photons were lost at $\delta r=30\arcsec$ and $60\arcsec$,
because of the assumed Gaussian shape used to derive analytically
extension and counts. We have applied a simple correction for the
object size to arrive at the good photometric results for \ewav\ 
presented on Fig.~\ref{test1:results}.

\section{ Test~2 -- point-like plus extended objects}
\label{sec:test2}

\subsection{Input configuration} 
This test is similar to Test~1, but we have replaced some of the
point-like sources by extended ones generated as described in
Sec.~\ref{sec:sim}. The raw photon image with input counts indicated
and its representations are shown in Fig.~\ref{test2:sum}.

\begin{figure*}
\vspace{14cm}
  \caption{
    Test~2. The raw X-ray photon image for 10 ks exposure time (upper
    left). As in Fig.~\ref{test1:sum} three cases of separations are
    indicated, as well as the corresponding input source counts. The
    extended objects are in the right columns. The Gaussian
    convolution with FWHM$=20\arcsec$ (upper right), \mr\ WT filtered
    image (lower left) and \wav\ reconstructed image both with a
    significance threshold of $10^{-4}$ (lower right) are shown.}
  \label{test2:sum}
\end{figure*}

\subsection{Detection rate and positional errors}

The number of missed and false detections are shown in
Tab.~\ref{test2:tab}.  An increase of the searching radius to
$20\arcsec$ was needed: at $\delta r=15\arcsec$ the blending tends to
shift the centroid towards the point-like source.  Note that this
situation is a clear case for source confusion: if we take the closest
neighbour (the point-like source in some cases) as the
cross-identification from the input list, we shall overestimate the
flux more than two-fold, while the true representation is the extended
object.

Some changes were needed for the procedures not based on the wavelet
technique in order to avoid splitting of the bright extended objects
into sub-objects: increase of the Gaussian convolution FWHM to
$20\arcsec$, and multi-PSF fit for \eml. In the Gaussian case, the
larger smoothing length smears some of the point-like sources, leading
to non-detection. \eml\ splitting persists even with the maximum
number of the PSFs fitted to the photon distribution (currently it is
capable of simultaneously fitting up to 6 PSFs).

\begin{table}
  \caption[]{
    Test~2 results for the detection rate. The first number in the
    ``Missed'' column is for point sources while the second is for
    extended ones. The total number of input sources is 24=12+12.}
  \label{test2:tab}
  \begin{tabular}{rcc}
    \hline\hline
    Method & Missed & False \\
    \hline
    \eml & 1+1 & 89\\
    \gsex $4\sigma$ & 6+0 & 1 \\
    \mrsex & 6+0 & 6 \\
    \wav   & 6+0 & 18 \\
    \ewav  & 4+2 & 5 \\
    \hline
  \end{tabular}
\end{table}

\subsection{Photometric reconstruction} 

\begin{figure*}
  \centerline{\hbox{
      \includegraphics[width=7cm]{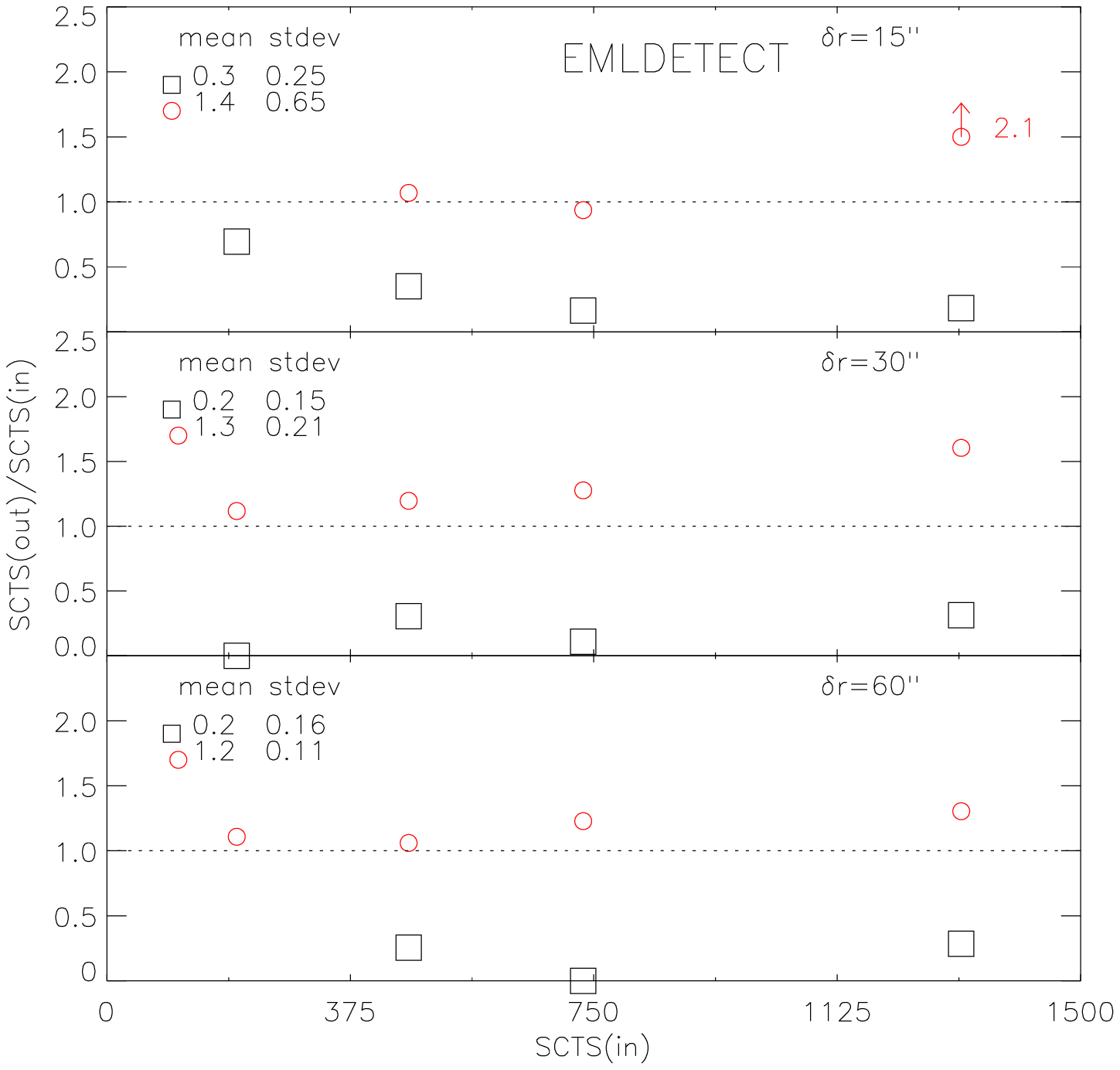} \hfil
      \includegraphics[width=7cm]{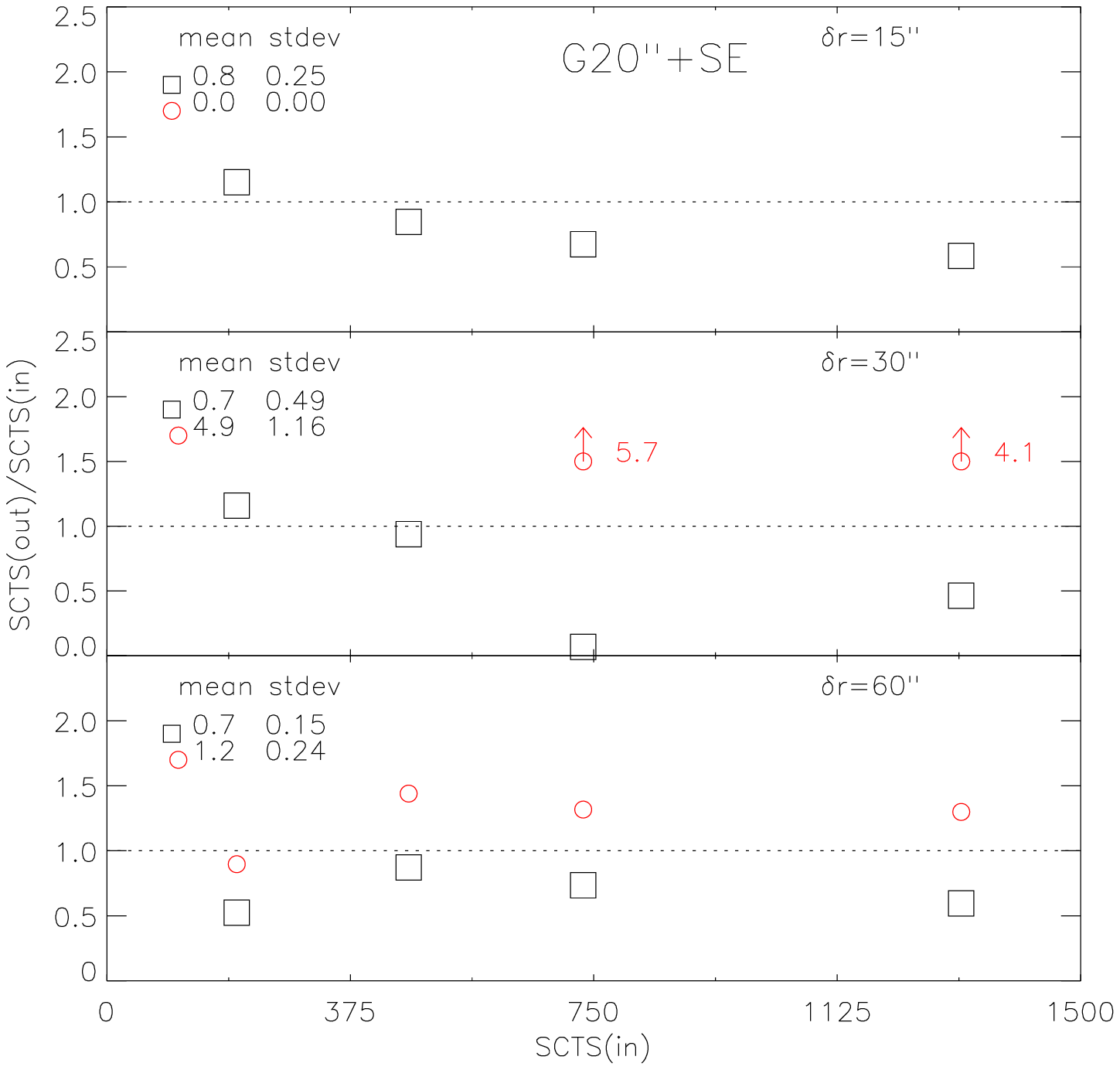}
      }}\medskip
  \centerline{\hbox{
      \includegraphics[width=7cm]{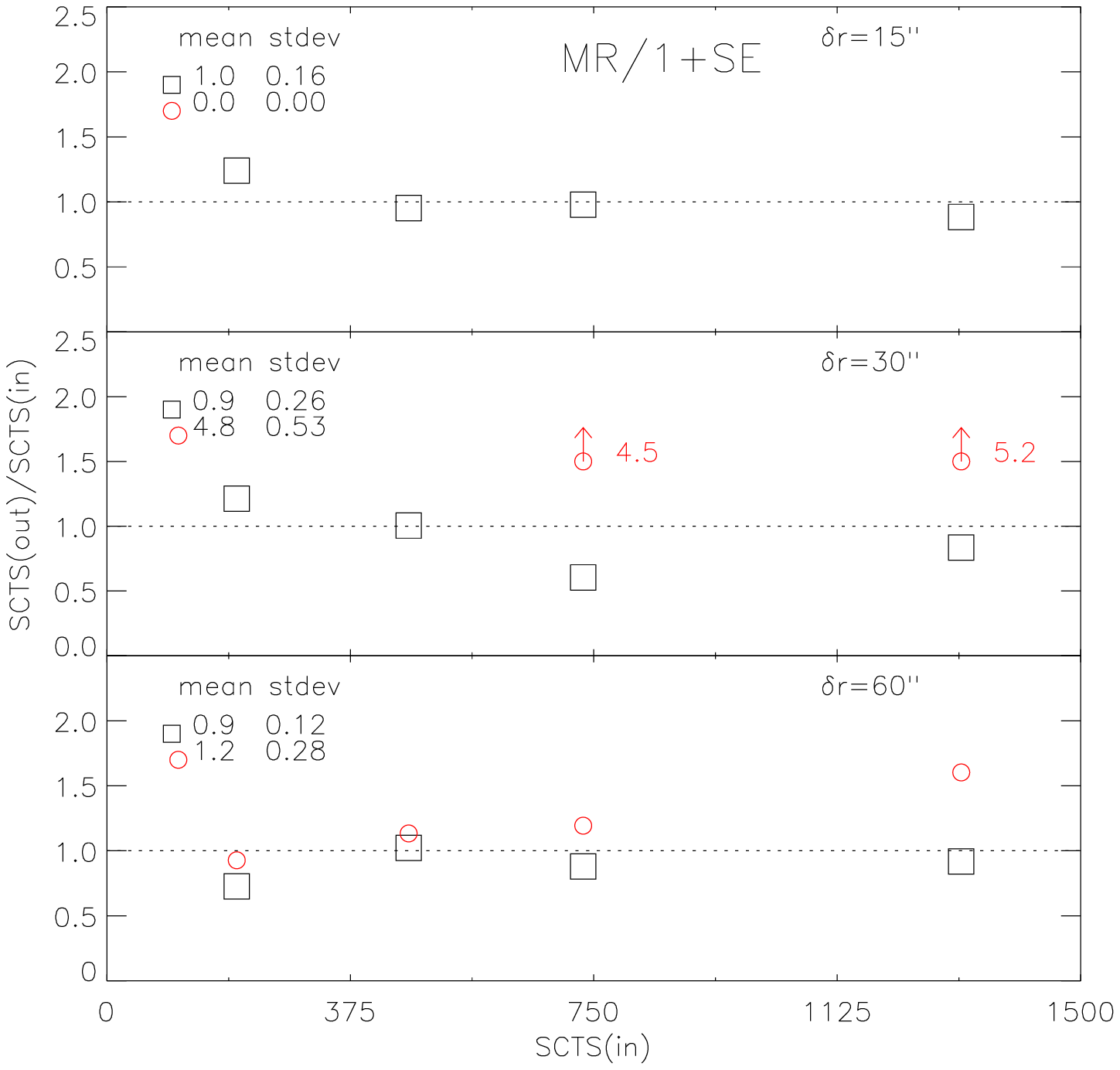} \hfil
      \includegraphics[width=7cm]{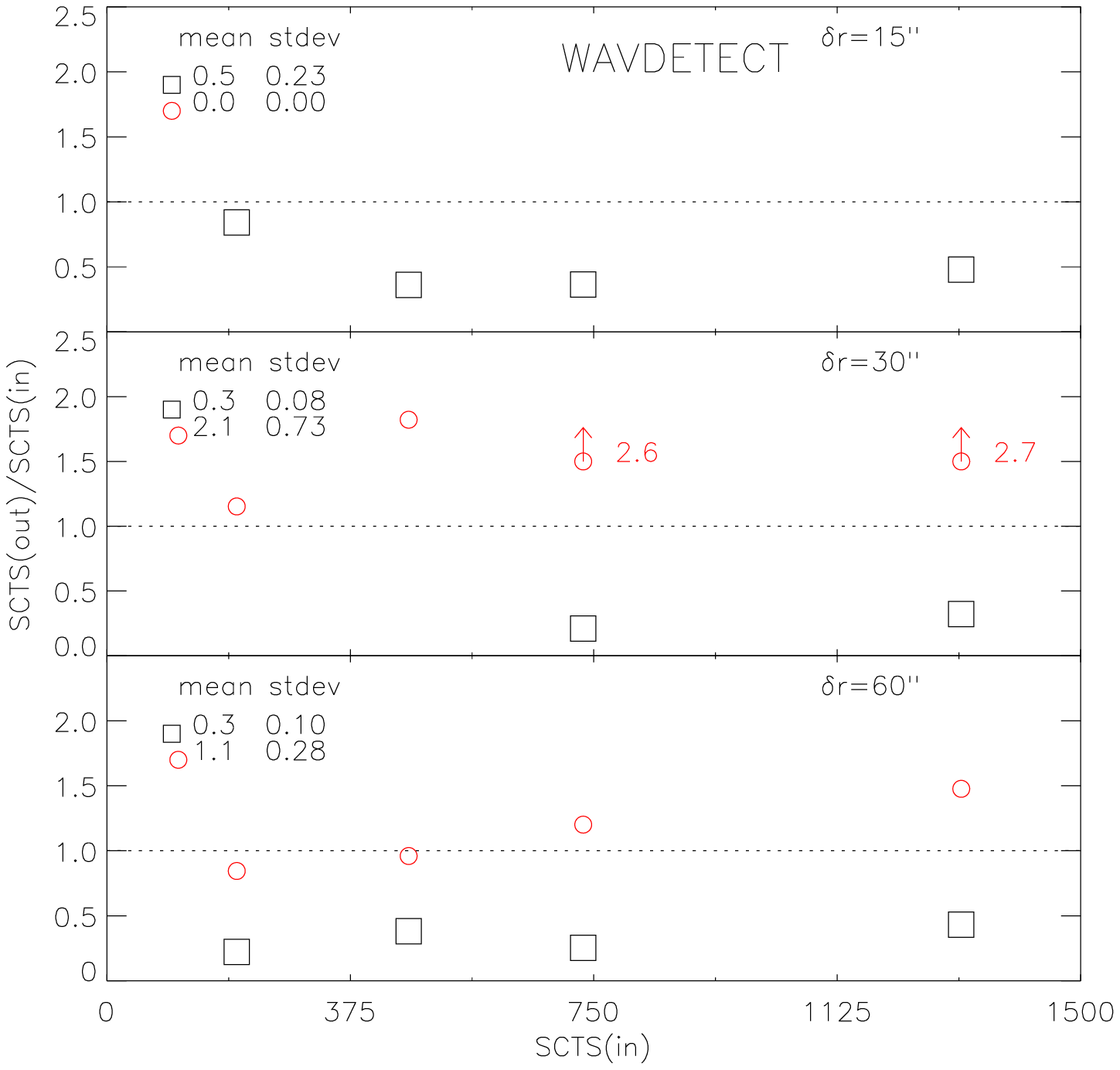}
      }}
  \caption{
    Test~2. As in Fig.~\ref{test1:results} except that the squares now
    represent extended objects and the point-like sources at fixed
    counts of 100 (circles) are put beside their corresponding
    neighbors (rather than being put at 100).  Circles with arrows and
    numbers denote the ratio when it is above 2.}
\label{test2:res}
\end{figure*}

The inferred-to-input source counts ratio $SCTS(out)/SCTS(in)$ is
shown in Fig.~\ref{test2:res}. We will not consider \ewav\  as its
photometric results and detection of extended sources were quite
unsatisfactory. The simple correction technique (as in Test~1) based
on the PSF does not work -- the extended objects profile has different
shape from the PSF.

\begin{description}
  
\item[$\delta r=15\arcsec$.] Only \mrsex\ gives good results for the
  flux restoration of the extended objects, but overestimating the
  counts for the faintest one. \wav\ misses almost half of the input
  photons while \eml\ splitting leads to very poor results.
  
\item[$\delta r=30\arcsec$.] All procedures give bad restoration
  results with \mrsex\ performing best again for the extended sources.
  The proximity of the objects leads to an overestimation of the
  point-like source counts and an underestimation of the extended
  object counts.  There is no simple way to correct for this effect,
  but can be done using a rather elaborated iterative procedure
  involving extended object profile fitting.
  
\item[$\delta r=60\arcsec$.] Point-like source results are relatively
  similar with all procedures -- the source counts are slightly
  overestimated due to the extended object halo even at $60\arcsec$.
  The problems of \wav\ and \eml\ and the recovery of the extended
  objects counts are quite obvious. Again \mrsex\ is the best
  performing procedure with extended objects flux uncertainty about
  25-30\%.

\end{description}

\subsection{Object classification} 

\begin{figure*}
  \centerline{\hbox{
      \includegraphics[width=7cm]{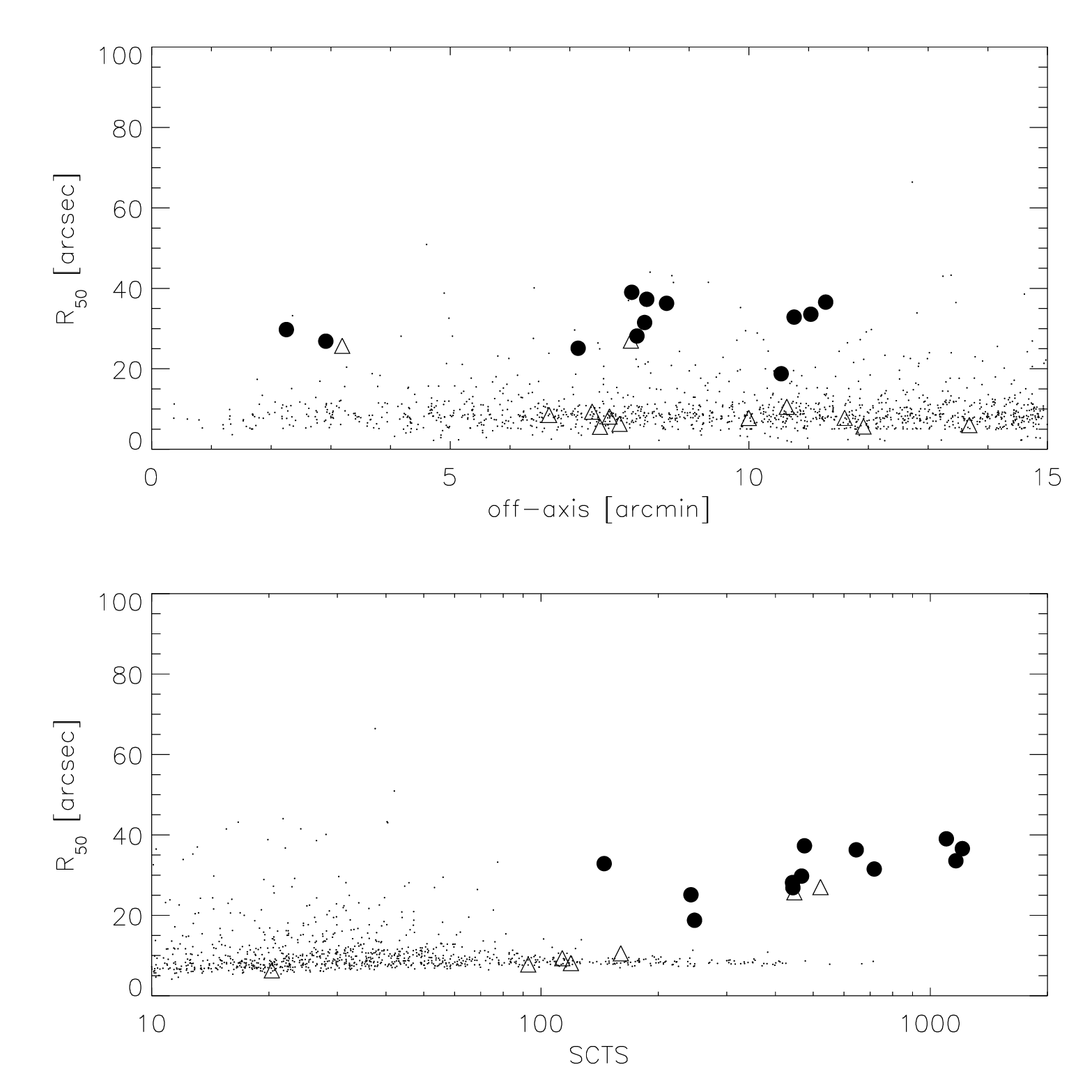} \hfil
      \includegraphics[width=7cm]{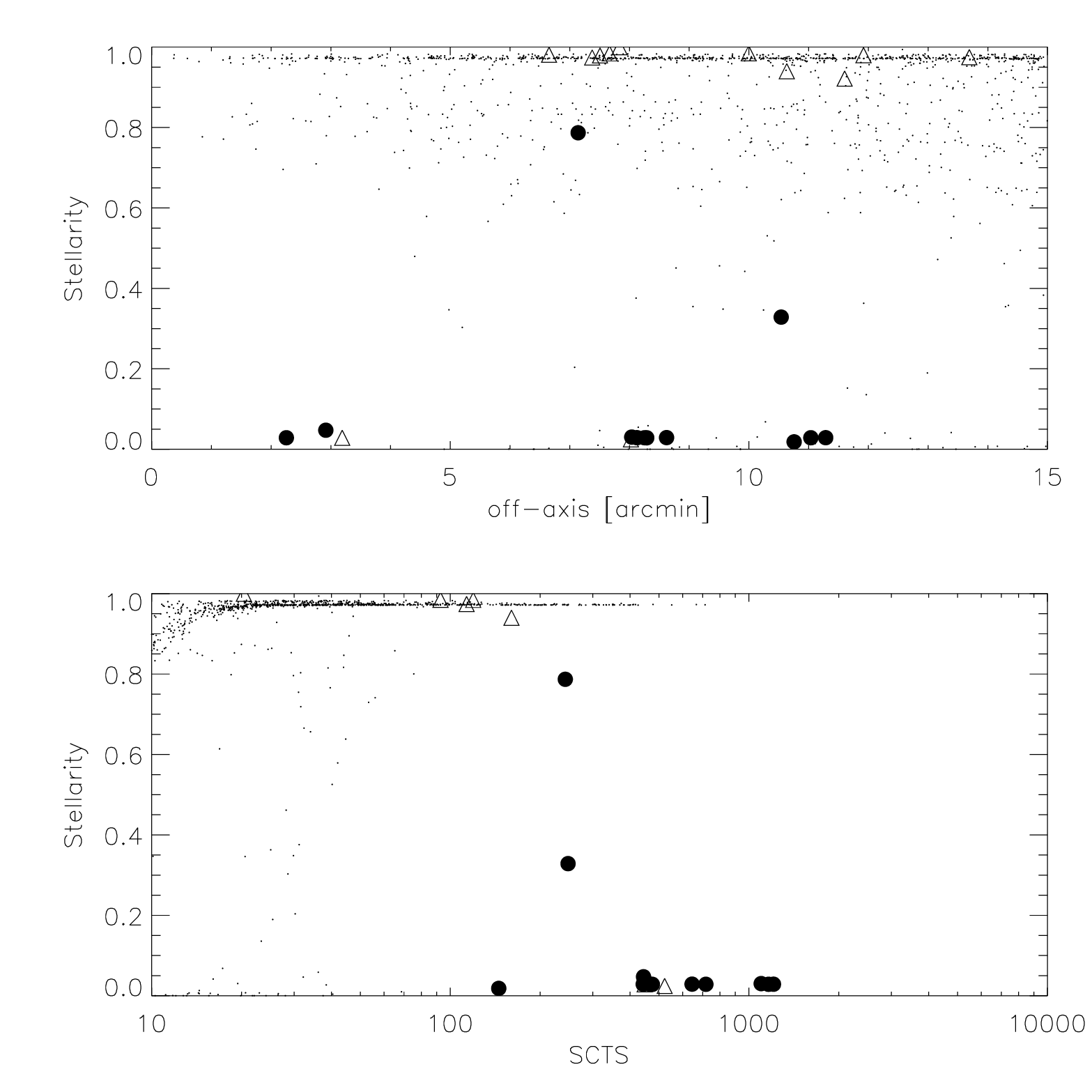}
      }}
  \caption{
    Test~2. \mrsex\  detection classification based on $R_{50}$ (left
    panels) and stellarity (right panels) as a function of the
    off-axis angle (upper panels) and detected source counts (lower
    panels). Identified extended (filled circles) and point-like
    objects (triangles) are plotted over the results from 10 simulated
    images with only point-like sources (see Sec.~\ref{sec:test3}).}
  \label{test2:class1}
\end{figure*}

An important test is the ability of the procedures to classify objects
and to allow further analysis of complicated cases of blending.  The
\mrsex results for the clasification by means of the half-light radius
and stellarity index (c.f.~Sec.~\ref{sec:class}) are shown in
Fig.~\ref{test2:class1}, overlayed over the results from 10 simulated
images with only point-like sources.  Clearly the detected extended
objects with \mrsex\ fall into zones not occupied by point-like
sources. Note, however, that the two detected point-like sources at
$30\arcsec$ will be mis-clasified as extended objects -- the proximity
not only influences the source counts, overestimated by more than 4-5
times (Fig.~\ref{test2:res}), but also the object profile and
consequently the classification.

\begin{figure*}
  \centerline{
    \includegraphics[width=7cm]{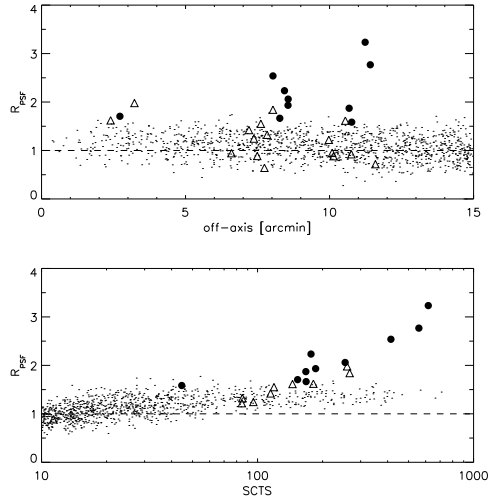}
    }
  \caption{
    Test~2. \wav\  classification based on object size to PSF size ratio
    -- $R_{PSF}$. The identified extended (filled circles) and
    point-like sources (triangles) are plotted over the results from
    10 simulated images with only point-like sources (see
    Sec.~\ref{sec:test3}), the dashed line marks a ratio of unity .}
  \label{test2:class2}
\end{figure*}

Fig.~\ref{test2:class2} shows the \wav classification -- the ratio of
the object size to the PSF size ($R_{PSF}$). The results are more
ambiguous with \wav\ (Fig.~\ref{test2:class2}) compared to \mrsex.
The results with \eml\ and its classification parameter (extension
likelihood) were very unsatisfactory due to the extended object
splitting.  More comprehensive discussion of the simulations and the
classification is left for Sec.~\ref{sec:test4}.

\subsection{Discussion}

Clearly \eml\ and \wav\ have problems in restoring the fluxes of
extended objects. We have already discussed the splitting difficulties
of \eml. The explanation for \wav's bad results is that the wavelet
scale at which the detected object size is closer to the PSF size
defines the source detection cell (in which the flux is computed). If
the characteristic size of an object is larger than the PSF size (i.e.
an extended object) this procedure will tend to underestimate the
flux.

We can safely accept the \mrsex\ procedure as the best performing for
detection and characterization both for point-like and extended
objects. We must stress however that one cannot rely on the flux
measurements when there are extended and point-like sources separated
by less than $30\arcsec$. The proximity affects also the
classification of the point-like sources. Using the classification and
then performing more complicated analysis like profile fitting and
integration for the extended sources can improve a lot the
restoration.  In realistic situations we can expect very often
problems of this kind, especially with \XMM.

\section{Test~3 -- Realistic model with only point-like sources}
\label{sec:test3}

\subsection{Input configuration}

We simulate an extragalactic field including only point-like sources
with fluxes drawn from the $\log N-\log S$ relation (Hasinger et al.
\cite{has98}, \cite{has01}, Giacconi et al. \cite{gia01}). PSF,
vignetting and background models are applied as described in
Sec.~\ref{sec:sim}. The aim is to test the detection procedures in
more realistic cases where confusion and blending effects are
important and not controlled. The raw photon image is shown in
Fig.~\ref{test3:sum} together with its visual representation -- the
same input configuration for a much larger exposure time and no
background, only keeping the objects with counts greater than 10. It
displays better the input object sizes, fluxes and positions and it is
instructive to compare it to the \mr\ filtered and \wav\ images shown
on the same figure.

\begin{figure*}
\vspace{14cm}
  \caption{
    Test~3. A simulated \XMM extragalactic field with only point-like
    sources for 10ks exposure time and the total sensitivity of the
    three EPIC instruments (upper left), its representation for much
    larger exposure time and no background (upper right). The \mrsex\ 
    filtering (lower left) and \wav\  images both with $10^{-4}$
    significance threshold (lower right) are shown.}
  \label{test3:sum}
\end{figure*}

\subsection{Cross-identification and positional error}

We need to define a searching radius in order to cross-identify the
output and the input lists. The input list contains many objects with
counts well below the detection limit ($\log N-\log S$ extends to very
faint fluxes) and a lower limit must be chosen.  For each detected
object, we search for the nearest neighbour inside a circle within the
reduced input list.

The positional difference for the brightest detected sources (more
than 100 counts) in the inner $10\arcmin$ from the center of the FOV
is shown in Tab.~\ref{tab:dr}.  The region beyond $10\arcmin$ is
subject to serious problems caused by the vignetting and PSF blurring,
the detected object centroid can be few PSFs widths from the true
input identification.

We therefore adopt the following cross-identification parameters: the
input list is constrained to counts greater than 10; a $6\arcsec$
searching radius; we consider only the central $10\arcmin$ of the FOV.

\begin{table}
\caption[]{One sigma positional error and number of detected
  objects inside the inner $10\arcmin$ from the center of the FOV and
  more than 100 counts for a 10 ks exposure.}
\label{tab:dr}
\begin{tabular}{rrr}
\hline
\hline
Procedure & $\Delta r\arcsec$ & number\\
\hline
\eml & 2.9 & 13 \\
\gsex & 3.5 & 14 \\
\mrsex & 3.2 & 13 \\
\wav & 4.1 & 12 \\
\hline
\end{tabular}
\end{table}

\subsection{Detection rate and photometric reconstruction}

\begin{figure*}[htb]
  \centerline{\vbox{\hbox{
        \includegraphics[width=7cm]{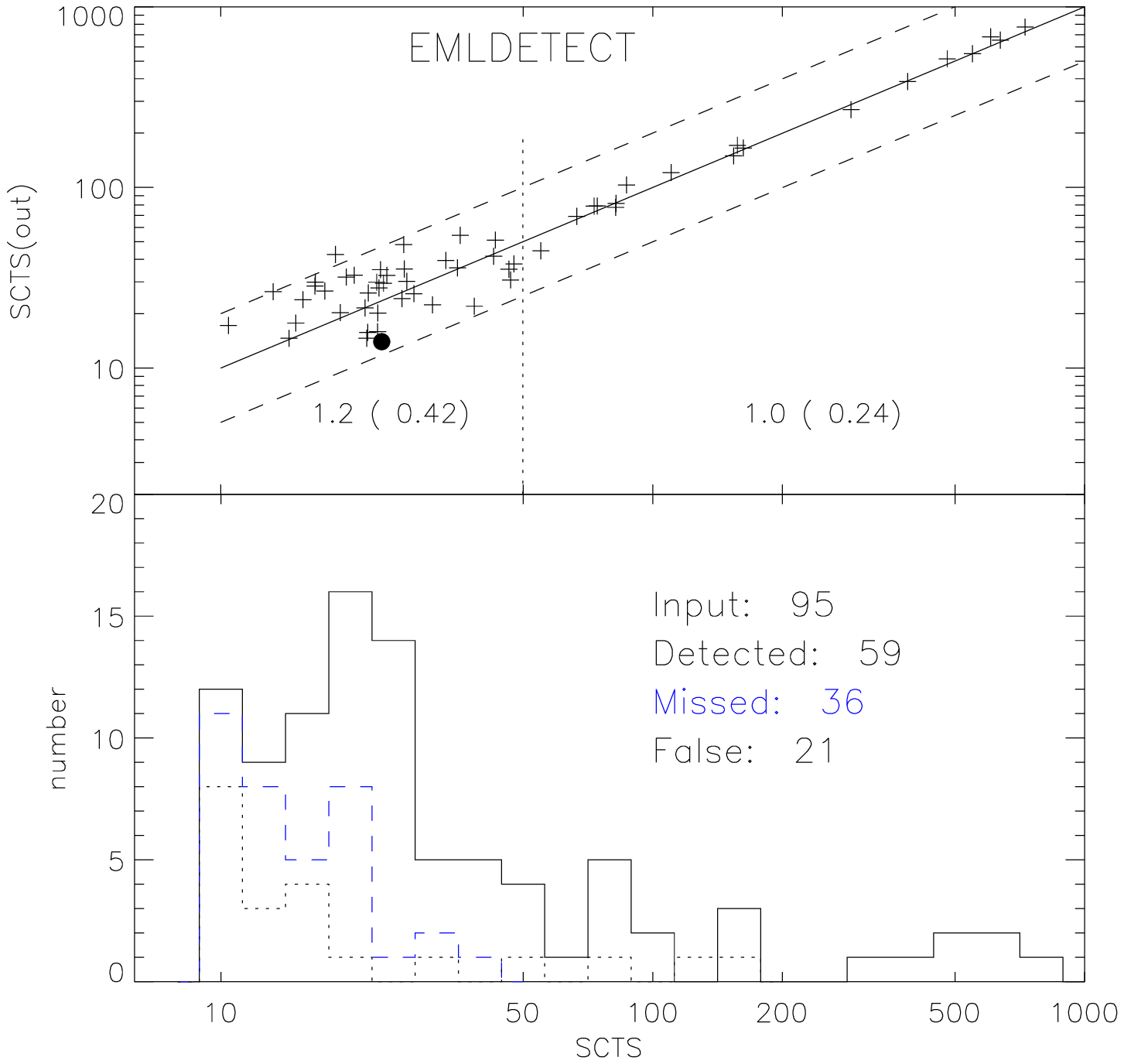} \hfil
        \includegraphics[width=7cm]{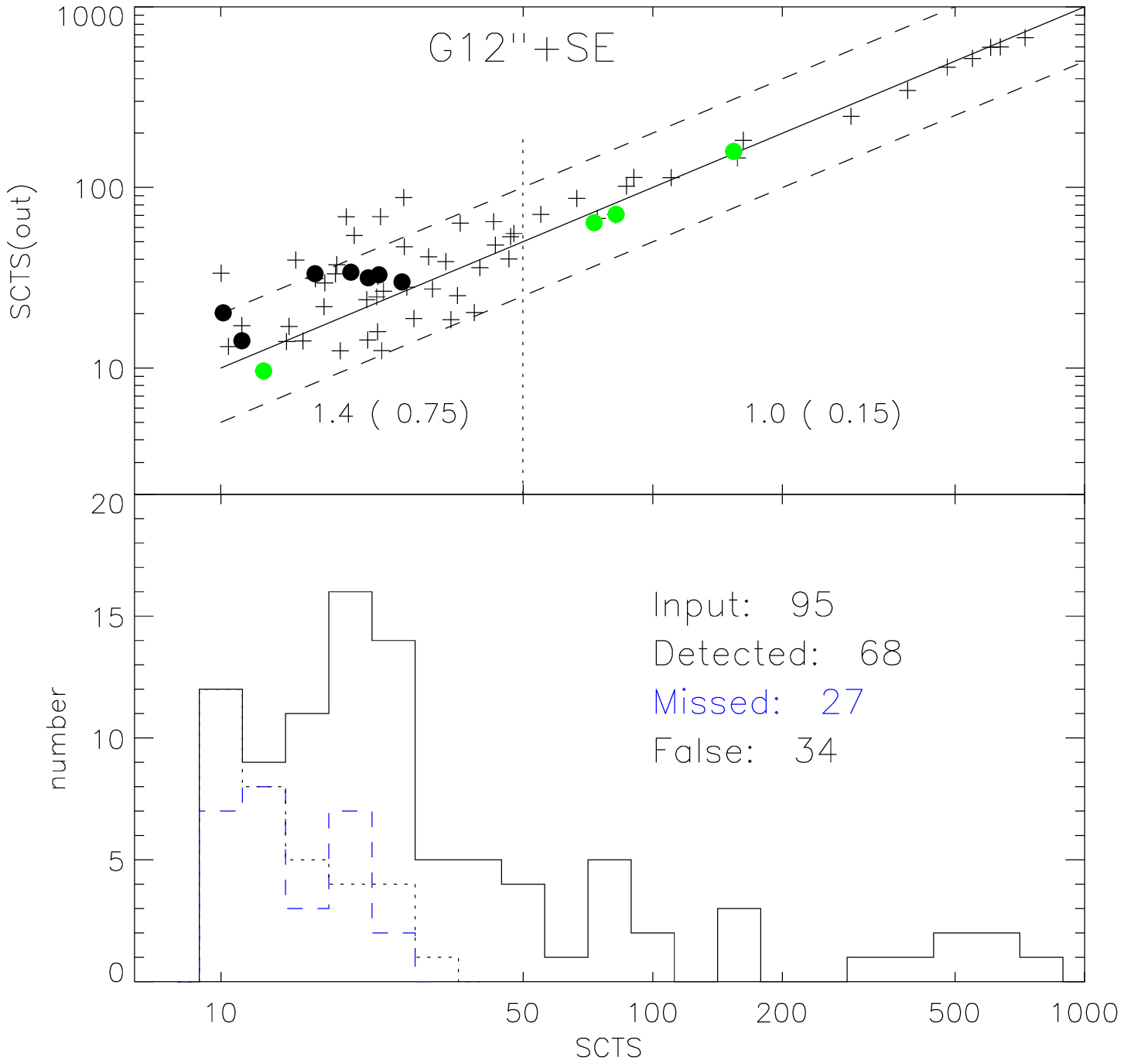}
        }\medskip
      \hbox{
        \includegraphics[width=7cm]{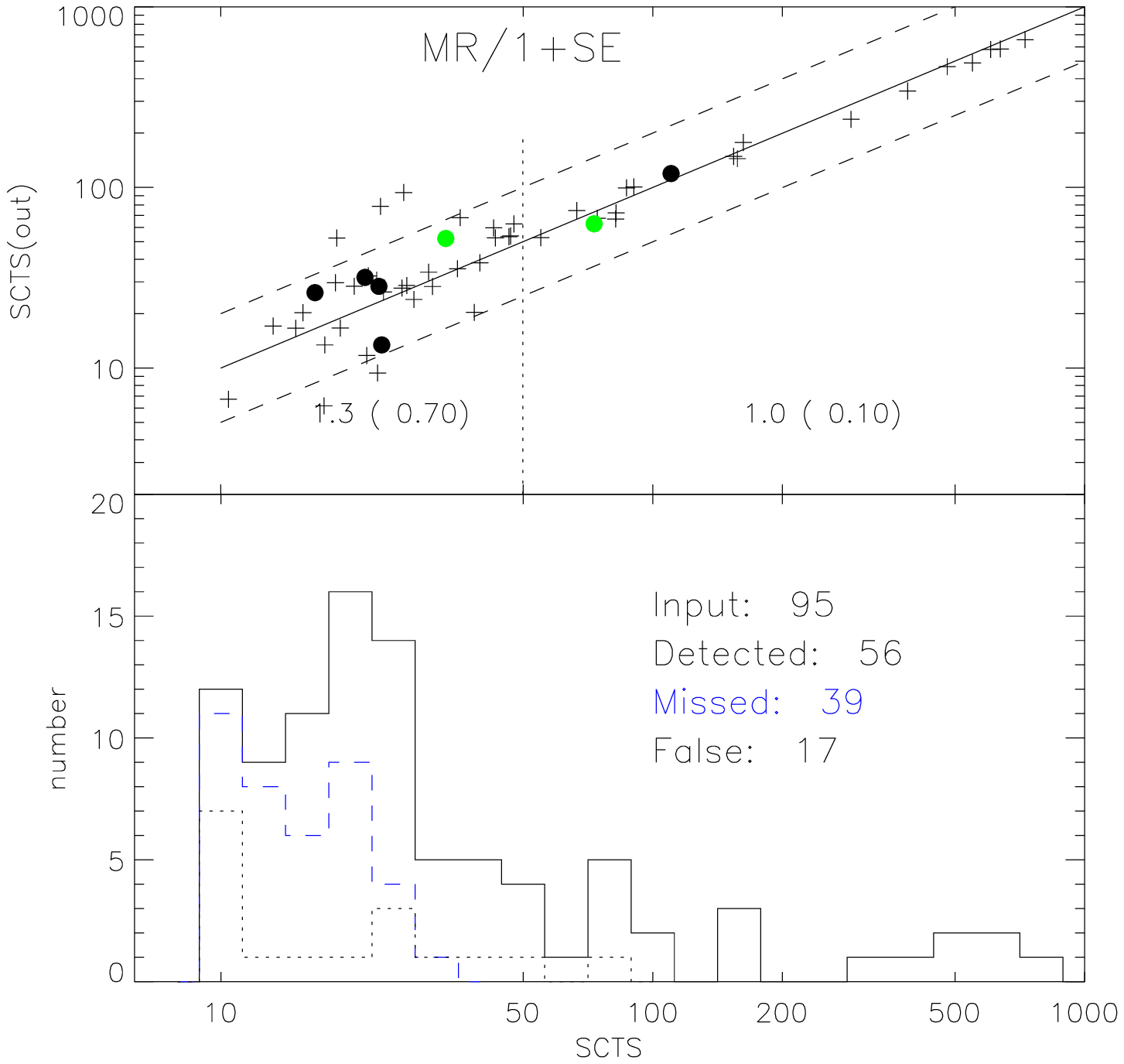} \hfil
        \includegraphics[width=7cm]{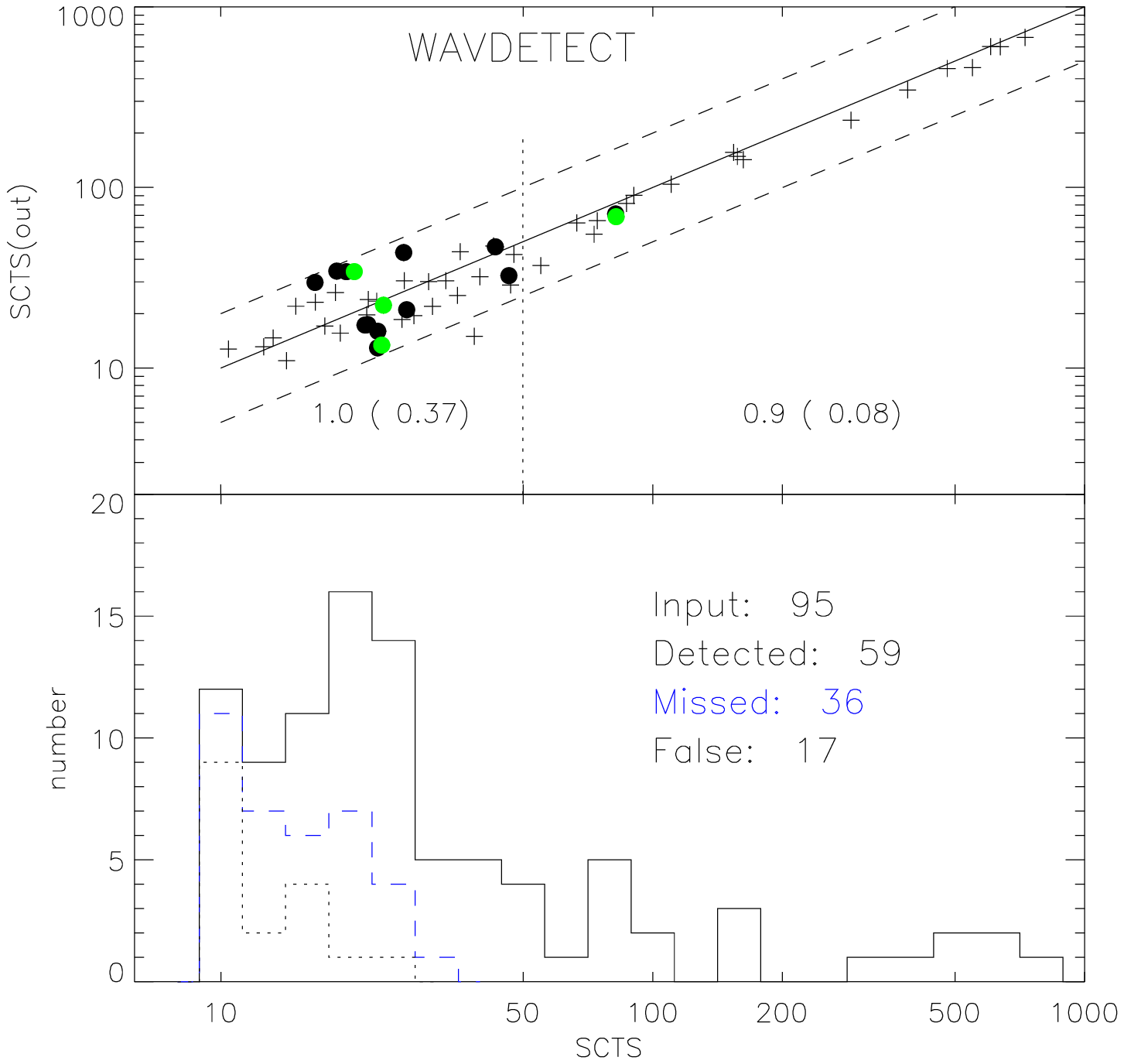}
        }}}
  \caption{
    Test~3. Recovery of the input flux (upper panel of each figure).
    The continuous line is exact match between detected and input
    counts while the dashed lines are for two times differences. The
    limit of 50 input counts is shown by a vertical dotted line and
    the mean and the st.dev. (in brackets) of $SCTS(out)/SCTS(in)$ for
    the two regions are indicated. The black circles denote objects
    with detect-input position difference larger than $4\arcsec$, 
    suggesting blending effects; the grey circles denote objects with
    more than one neighbour inside the searching radius. In the bottom
    panels, the corresponding rate and distribution of input counts
    (continuous histogram), missed input objects (dashed histogram)
    and possible false detections (dotted histogram) are shown.}
\label{test3:flux}
\end{figure*}

The detection rate and flux reconstruction results are shown in
Fig.~\ref{test3:flux}. There are different effects playing a role in
the distribution and the numbers of missed and false detections:
\begin{description}
\item[(1)] ``false'' detections -- non-existent objects, or two or
  more sources blended into a single detected object. The result will
  be a ``false'' detection if the blended objecs are not in the input
  list (count limit) or the merged object centroid is beyond the
  searching radius.
\item[(2)] source confusion -- in the cross-identification process the
  nearest neighbour to the detected source is not the true assignment;
  or as in case (1), when a blend of objects is wrongly identified by
  one input source.
\item[(3)] missed detections -- depending on the local noise
  properties, some objects can be missed even if their input counts are
  above the adopted limiting counts for cross-identification.
\end{description}

\subsection{Discussion}

The results in terms of detection rate are similar for all procedures.
The best detection rate shows \gsex\ but at the price of twice as many
false detections.

The photometry reconstruction for the sources above 50 counts shows a
spread about 10-15\% for the WT based methods and $\sim 25\%$ for
\eml. However, \eml\ clearly outperforms the other procedures when we
use the same PSF model as the one hard-coded into the programme. This
fact shows that using a correct PSF representation has a crucial
importance for the ML technique. More discussion about the detection
limits, completeness and confusion is left for Sec.~\ref{sec:test5}.

\section{Test~4 -- sky models with point-like and extended objects}
\label{sec:test4}

\subsection{Input configuration}
We have chosen exactly the same input point-like sources configuration
as in Test~3 and we have added 5 extended objects. They may be
regarded as clusters of galaxies at different redshifts with the same
$\beta$-profiles and moderate X-ray luminosity $L_{[2-10]keV} \sim
3\times 10^{44}$ erg/s (c.f. Tab.~\ref{tab:counts}). Now the objective
is to estimate the capabilities of the procedures to detect and
identify extended objects. The difference from Test~2 is the arbitrary
positions of the point-like sources leading to different local and
global background properties and to uncontrolled confusion effects.

The input configuration and wavelet filtered and output images are
given in Fig.~\ref{test4:sum}.

\begin{figure*}
\vspace{14cm}
  \caption{
    Test~4. The raw X-ray photon image with point-like and extended
    objects for 10ks exposure time (upper left) and its visual
    representation for much larger exposure and no background (upper
    right). The corresponding redshifts for the clusters are
    indicated. The \mrsex\  filtering (lower left) and \wav\  both with
    $10^{-4}$ significance threshold (lower right) are shown.}
  \label{test4:sum}
\end{figure*}

\subsection{Positional and photometric reconstruction}
Results for the point-like sources will not be presented, because the
input configuration (position, $\log N- \log S$, background) is
exactly the same as in the previous test.  It was shown in Test~2 that
the presence of a point source even with moderate counts in the
vicinity of a faint extended source could lead to confusion and even
non-detection. Of course, the presence of a cluster will affect the
detection and photometry of the point-like objects in its vicinity,
but this effect is of minor concern for this test and has been already
discussed (Test~2).

The detection rate, input-detect position offsets, detected counts and
detected-to-input counts ratio are shown in Tab.~\ref{tab:test4}.

\begin{table}
\caption[]{
  Recovery of position and flux of the extended objects.}
\label{tab:test4}
\begin{tabular}{rrrrr}
\hline\hline
\multicolumn{1}{c}{redshift} & 
\multicolumn{1}{c}{$\Delta r$} & 
\multicolumn{1}{c}{Input} & 
\multicolumn{1}{c}{Detect} & 
\multicolumn{1}{c}{Detect/Input} \\
& \multicolumn{1}{c}{[arcsec]} & 
 \multicolumn{1}{c}{[counts]} & 
 \multicolumn{1}{c}{[counts]} & 
 \multicolumn{1}{c}{[\%]} \\
\hline
\multicolumn{5}{c}{\eml} \\
 0.6 &    2.1 &  1316 &    94 &    7 \\
 1.0 &    8.0 &   465 &    12 &    3 \\
 1.5 &   12.3 &   200 &   161 &   81 \\
 1.8 &    4.6 &   136 &   228 &  167 \\
 2.0 &   14.8 &   109 &    32 &   29 \\
\hline
\multicolumn{5}{c}{\gsex}\\
 0.6 &    1.2 &  1316 &  1043 &   79 \\
 1.0 &    5.2 &   465 &   355 &   76 \\
 1.5 &    1.9 &   200 &   220 &  110 \\
 1.8 & \multicolumn{4}{c}{Not detected}\\
 2.0 &   15.3 &   109 &    80 &  73 \\
\hline
\multicolumn{5}{c}{\mrsex}\\
 0.6 &    0.2 &  1316 &  1016 &   77 \\
 1.0 &    2.3 &   465 &   340 &   73 \\
 1.5 &    1.8 &   200 &   223 &  111 \\
 1.8 &   11.8 &   136 &    83 &   61 \\
 2.0 &   10.8 &   109 &   185 &  169 \\
\hline
\multicolumn{5}{c}{\wav}\\
 0.6 &    5.8 &  1316 &   344 &   26 \\
 1.0 &   10.6 &   465 &   193 &   41 \\
 1.5 &    0.1 &   200 &    39 &   19 \\
 1.8 & \multicolumn{4}{c}{Not detected}\\
 2.0 &   15.3 &   109 &    27 &   24 \\
\hline
\end{tabular}
\end{table}

As to the positional errors, it was already concluded that the centres
of the extended object can be displaced by more than the adopted
searching radius for point-like sources (Sec.~\ref{sec:test2}). The
differences in positions shown in Tab.~\ref{tab:test4}, especially for
fainter objects, are 3-4 times larger than the one sigma limit for
point-sources inside the inner $10\arcmin$ of the FOV
(Tab.~\ref{tab:dr}).

It is confirmed again that \eml\ and \wav\ are not quite successful in
charactering extended objects. But note that this time the results for
\mrsex\ and \gsex\ are worse than the results in Test~2 -- the rate of
lost photons being about 20-30\%. Also, the flux of the clusters at
$z=1.5$ and $2$ is overestimated, suggesting blending with faint
nearby point-like sources.

\subsection{Classification}
To classify the detected objects we have performed many simulations
with only point-like sources as in Test~3. Ten simulated images were
generated with the same $\log N-\log S$ and background, but with
different and arbitrary positions of the input sources. Exactly the
same parameters were used for filtering, detection and
characterization.

\begin{figure*}[htb]
  \centerline{\hbox{
      \includegraphics[width=7cm]{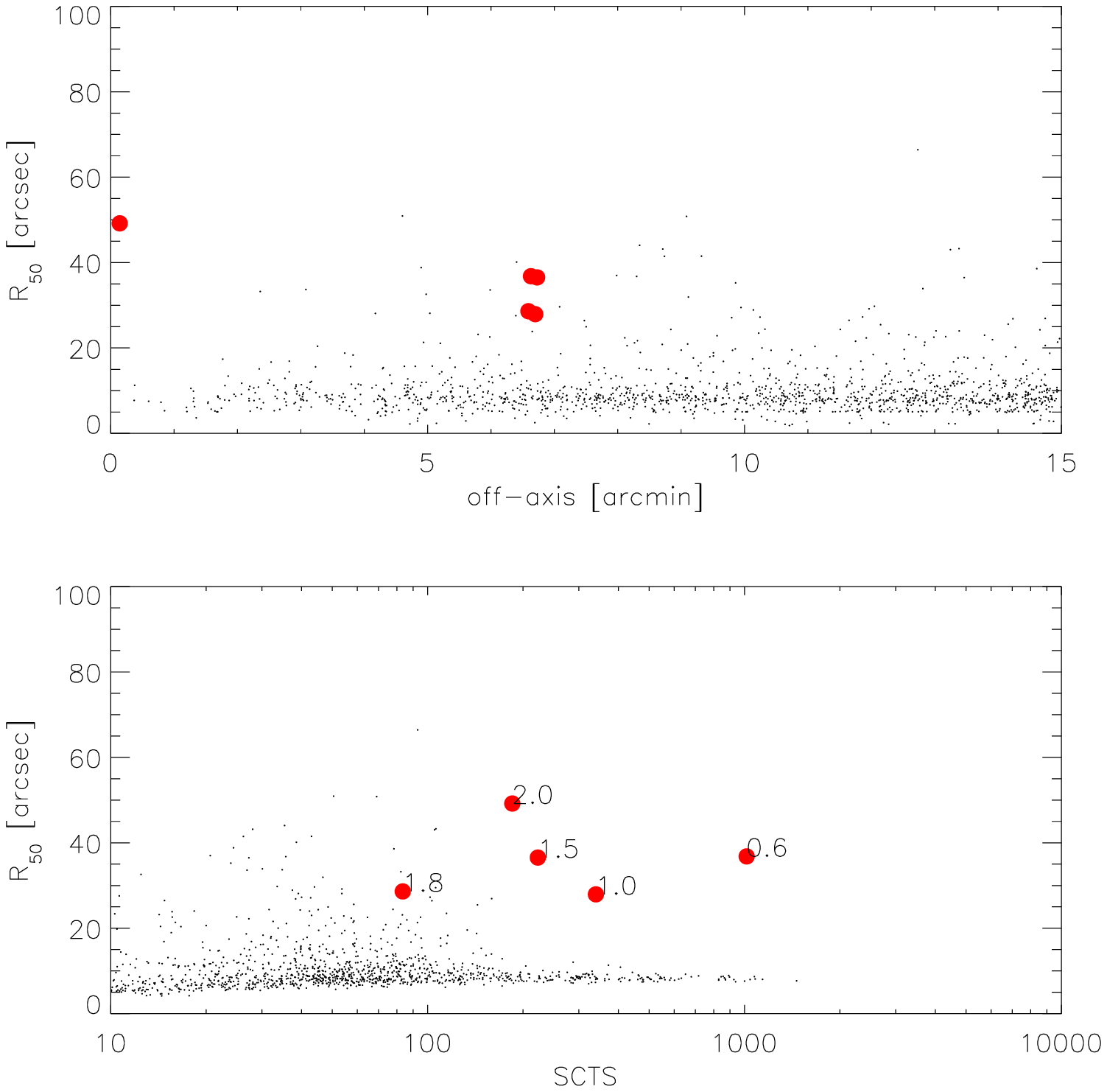} \hfil
      \includegraphics[width=7cm]{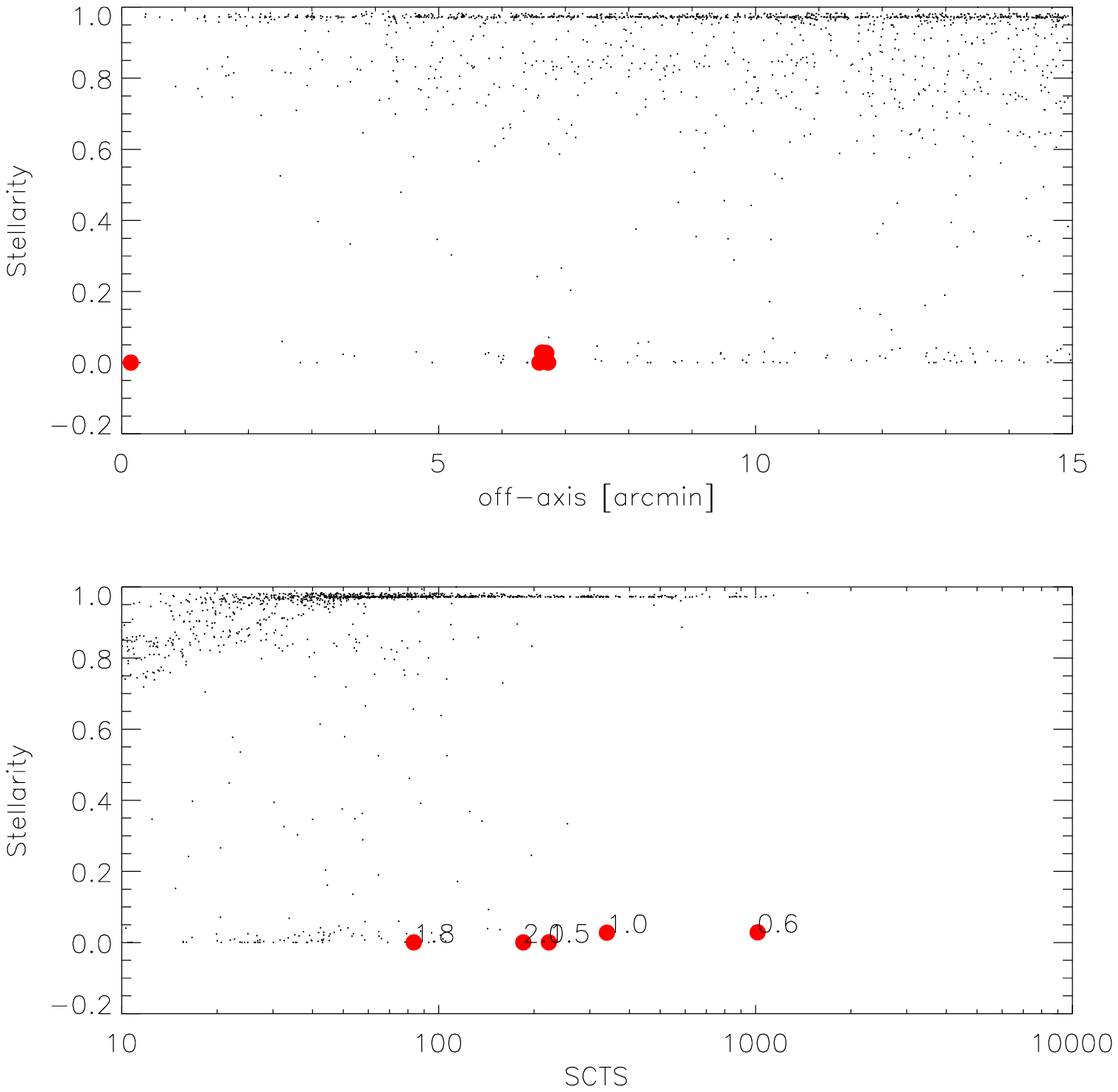}
      }}
  \caption{
    Test~4. Half-light radius $R_{50}$ (left figures) and stellarity
    index (right) as a function of the off-axis angle (up) and
    detected counts (down) for 10 generations.  The detection was
    performed with \sex\  after \mr\  multiresolution filtering. There are
    in total 1287 detections indicated by points.  The extended
    objects from this test are shown with filled circles.}
\label{test4:class}
\end{figure*}

Two classification parameters were used: the half-light radius
($R_{50}$) and the stellarity index from \sex\ based procedures. The
results are shown in Fig.~\ref{test4:class}.  We do not show results
with \wav\ and \eml\ classification parameters because their
unsatisfactory results were confirmed, as in Test~2.

We can see the excellent classification based on the stellarity index
and half-light radius: in the inner $10\arcmin$, for objects with more
than 20 detected counts, stellarity less than 0.1 and $R_{50} \geq
20\arcsec$ we have 15 incorrect assignments from 1287 detections
($\sim 1\%$).

\section{Test~5 -- completeness and confusion}
\label{sec:test5}

In this section we investigate the confusion and completeness problems
for \XMM shallow and deep observations like the first XMM Deep X-ray
Survey in the Lockman Hole (Hasinger et al. \cite{has01}).

A set of 10 images with exposure times of 10 ks and 100 ks in the
energy bands [0.5-2] and [2-10] keV were generated; the fluxes were
drawn using the latest $\log N - \log S$ relations from Hasinger et
al.  (\cite{has01}) and Giacconi et al. (\cite{gia01}).  Detection and
analysis were performed with exactly the same parameters for all
simulations: detection threshold, analysis threshold, background map
size, detection likelihood, etc. (see Sec.~\ref{sec:det}).
Cross-identification was achieved using the input sources above 10
counts and 30 counts for 10 ks and 100 ks exposures respectively.
Lowering the count limits yields more cross-identifications but
increases considerably the number of spurious detections.

The input image for 100 ks and [0.5-2] keV band is shown in
Fig.~\ref{test5:fig2}. The inner $10\arcmin$ zone where all analysis
is performed is indicated, as well as the total \XMM field-of-view.
It is informative to compare it with the images for 10 ks in
Fig.~\ref{test3:sum}.

\begin{figure*}
\vspace{7cm}
  \caption{Simulated 100ks \XMM deep field in [0.5-2] keV with the same
    parameters ($N_H,\ \log N - \log S,\ \Gamma$, background) as in
    the Lockman Hole (Hasinger et al.  \cite{has01}, Watson et al.
    \cite{wat00}). We restrict the analysis to the inner $10\arcmin$.
    }
  \label{test5:fig2}
\end{figure*}

In order to estimate the effect of confusion we have generated images
with only point-like sources, distributed on a grid such to avoid
close neighbours, and with fixed fluxes spanning the interval
$[10^{-16},10^{-13}]$ \flux.

In the following we discuss some important points.
\begin{itemize}
\item {\it Confusion and completeness}\\
  The detection rate of the input sources as a function of flux
  (Fig.~\ref{test5:det}) indicates that confusion problems are
  significant for 100 ks exposures in the [0.5-2] keV band. They are
  marginal for 100 ks exposures in [2-10] keV band and absent for 10
  ks in both bands.
  
  \begin{figure*}
    \centerline{\hbox{
        \includegraphics[width=7cm]{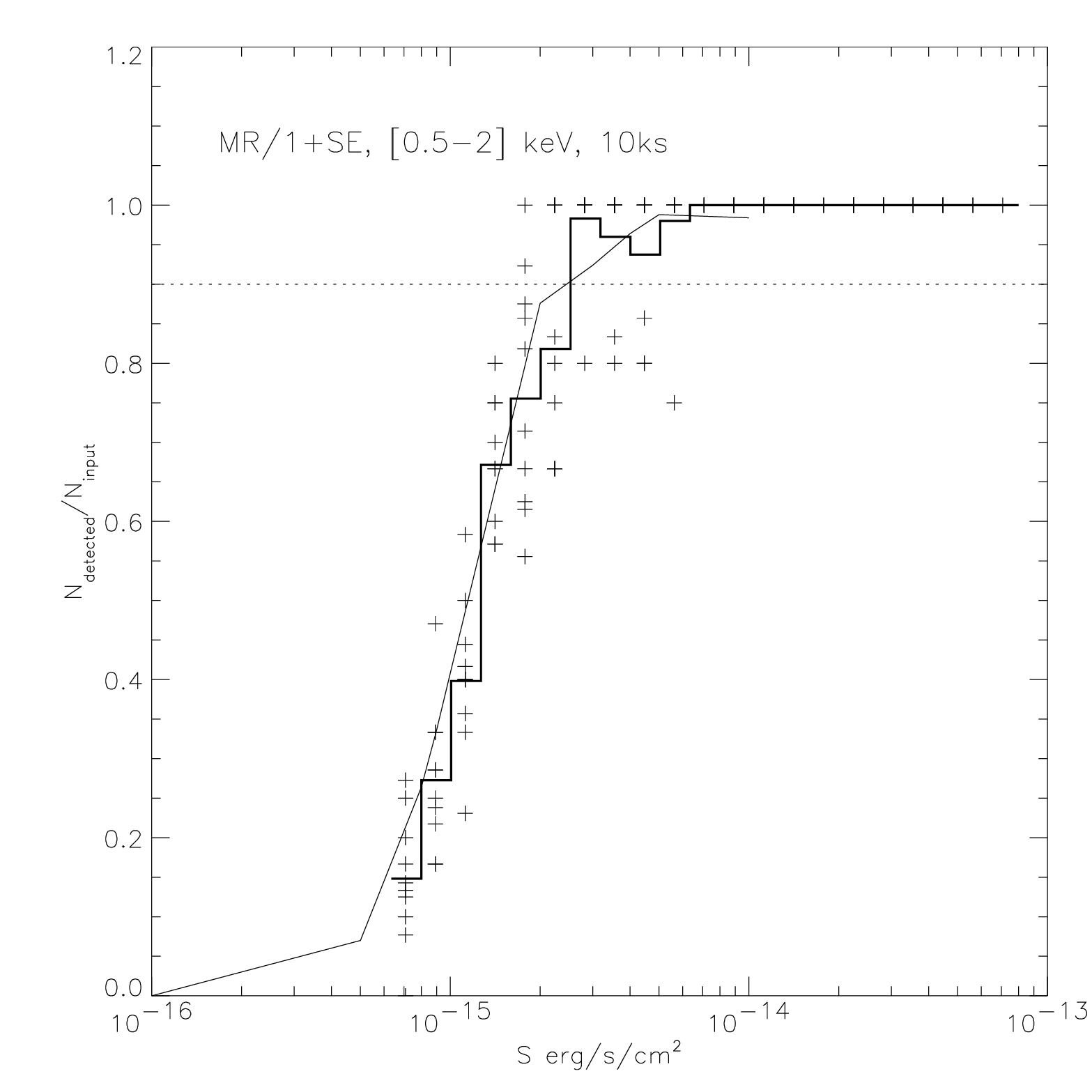} \hfill
        \includegraphics[width=7cm]{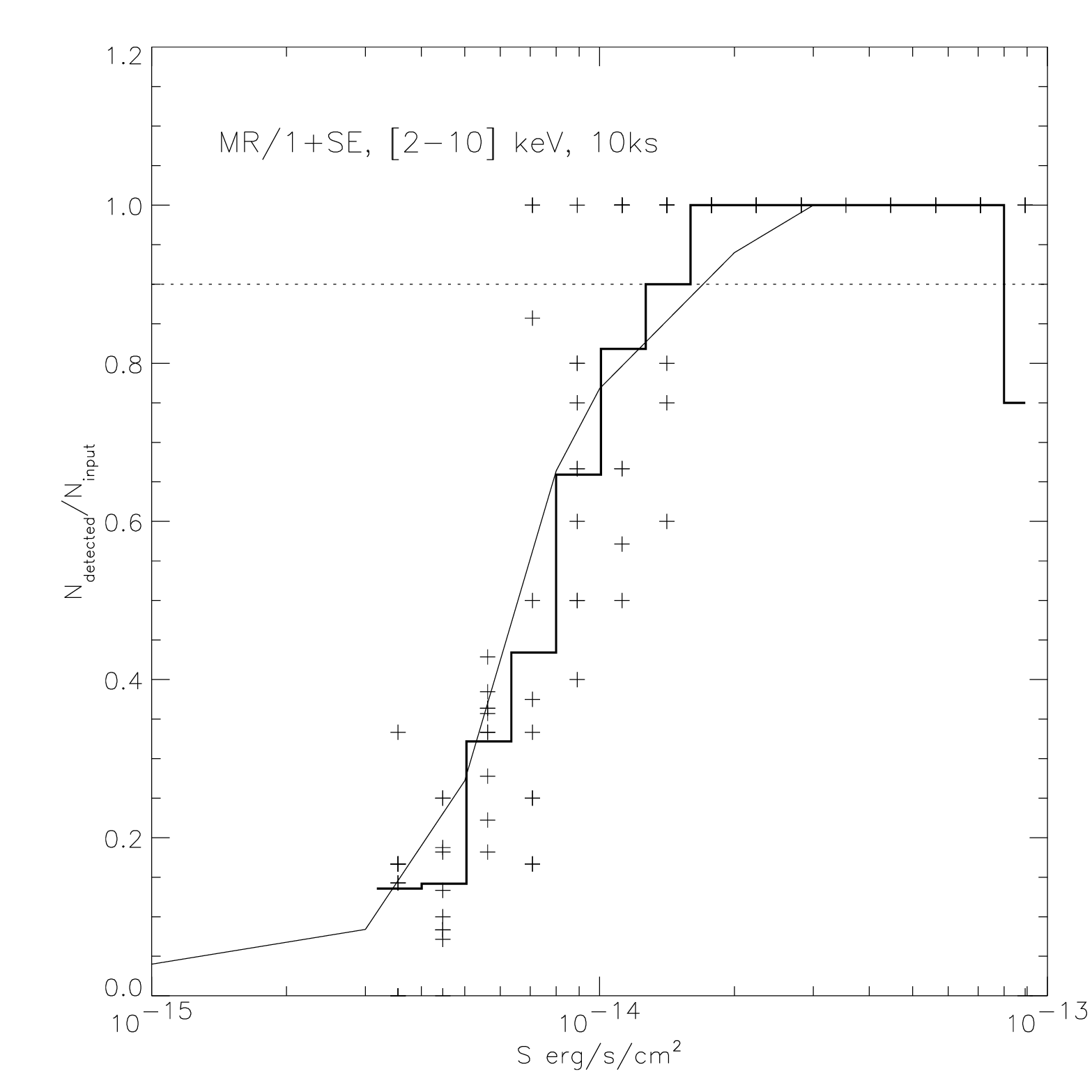}
        }}\medskip
    \centerline{\hbox{
        \includegraphics[width=7cm]{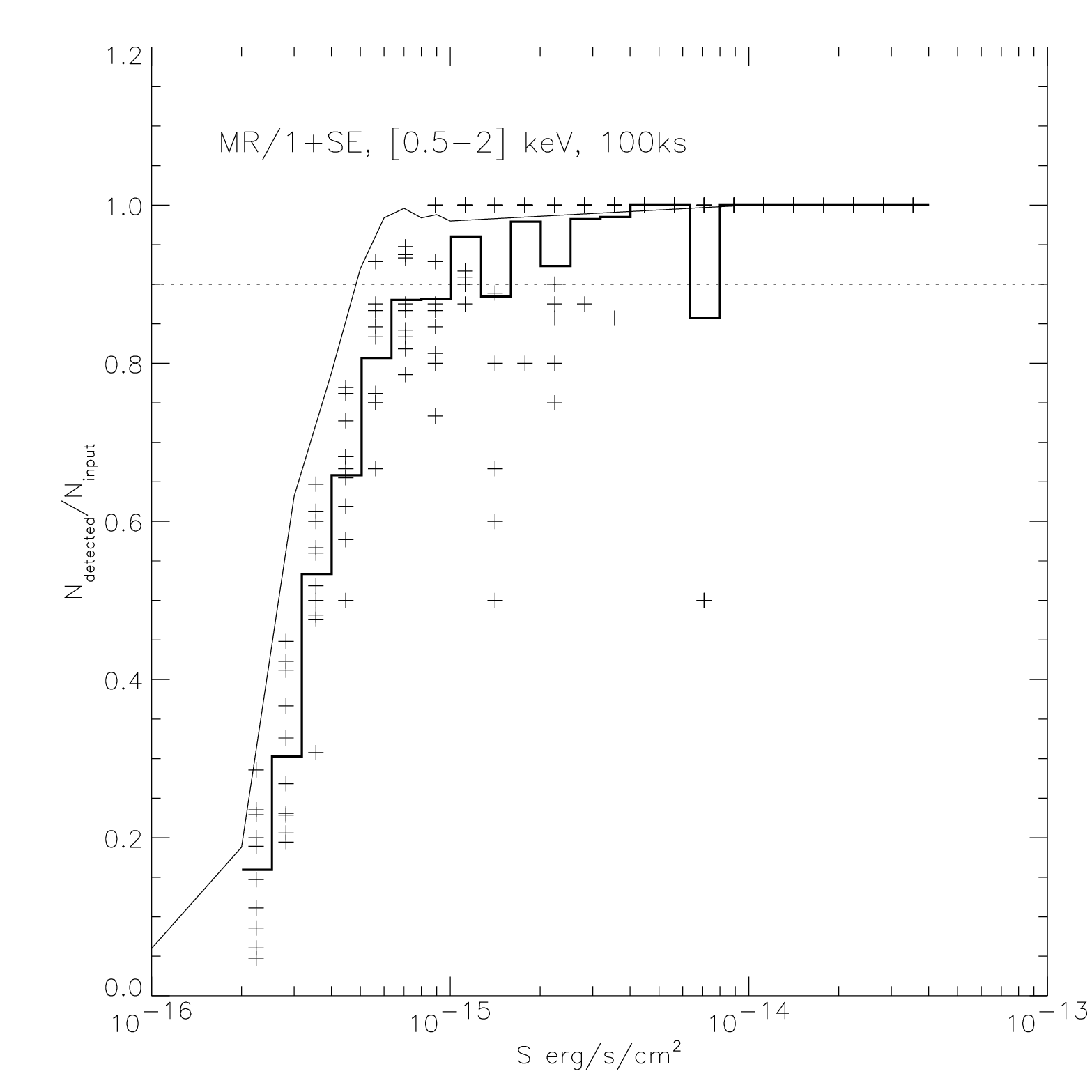} \hfill
        \includegraphics[width=7cm]{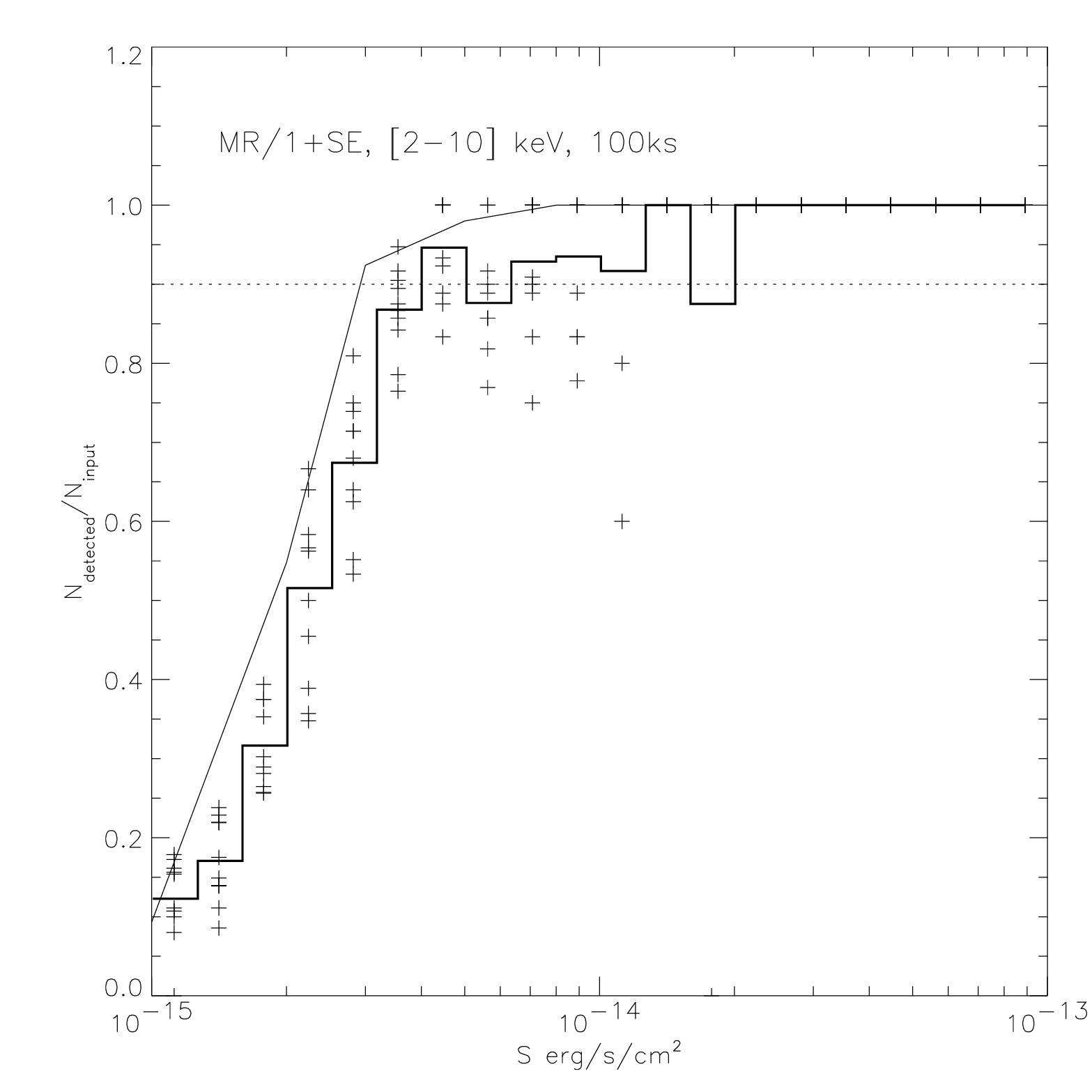}
        }}
    \caption{
      Ratio of number of cross-identified objects to the number of
      input objects, with counts greater than 10 and 30 for 10 ks and
      100 ks exposures respectively, as a function of the input flux.
      The results of 10 generations are shown by crosses, the heavy
      histogram is the corresponding average; the curve indicates the
      detection rate if confusion is absent and the dotted line marks
      90\% completeness.}
    \label{test5:det}
  \end{figure*}
  
  Energy conversion factors and the limiting fluxes below which more
  than 10\% of the input objects are lost are shown in
  Tab.~\ref{tab:test5}.
  
  \begin{table}
    \caption[]{
      The energy conversion factors (ECF) and the 90\% completeness limits
      for detections. ECF is computed assuming power-low spectrum with
      photon index $\Gamma=2,\ N_H=5\times 10^{19}$ cm$^{-2}$
      (Hasinger et al. \cite{has01}) and the
      three EPIC instruments (pn+2MOS) with thin filters.}
    \label{tab:test5}
    \begin{tabular}{rrr}
      \hline
      \hline
      & [0.5-2] keV     & [2-10] keV \\
      \hline
      \multicolumn{3}{c}{ECF (cts/s per \flux)}\\
      & $6.70\times 10^{-13}$ & $3.66\times 10^{-12}$ \\
      \multicolumn{3}{c}{Flux limits (\flux)}\\
      10ks  & $2\times 10^{-15}$ & $ 10^{-14}$ \\
      100ks & $6\times 10^{-16}$         & $3 \times 10^{-15}$ \\
      \hline
    \end{tabular}
  \end{table}
  
\item {\it Differential flux distribution}\\
  The differential flux distributions for 100 ks in [0.5-2] keV
  obtained by \mrsex\ and \eml\ are shown on Fig.~\ref{test5:diff}.
  Spurious detections appears to be numerous with \eml\ below $5\times
  10^{-16}$ \flux\ and tend to compensate the loss of sensitivity and
  confusion. \mrsex\ appears to be less affected and allows us to set
  a conservative flux limit of $6\times 10^{-16}$ \flux ($\sim 90$
  photons for 100 ks), below which the incompleteness becomes
  important -- 65\% of the input sources are lost between $3\times
  10^{-16}$ and $6\times 10^{-16}$ \flux.

  \begin{figure*}
    \centerline{\hbox{
        \includegraphics[width=7cm]{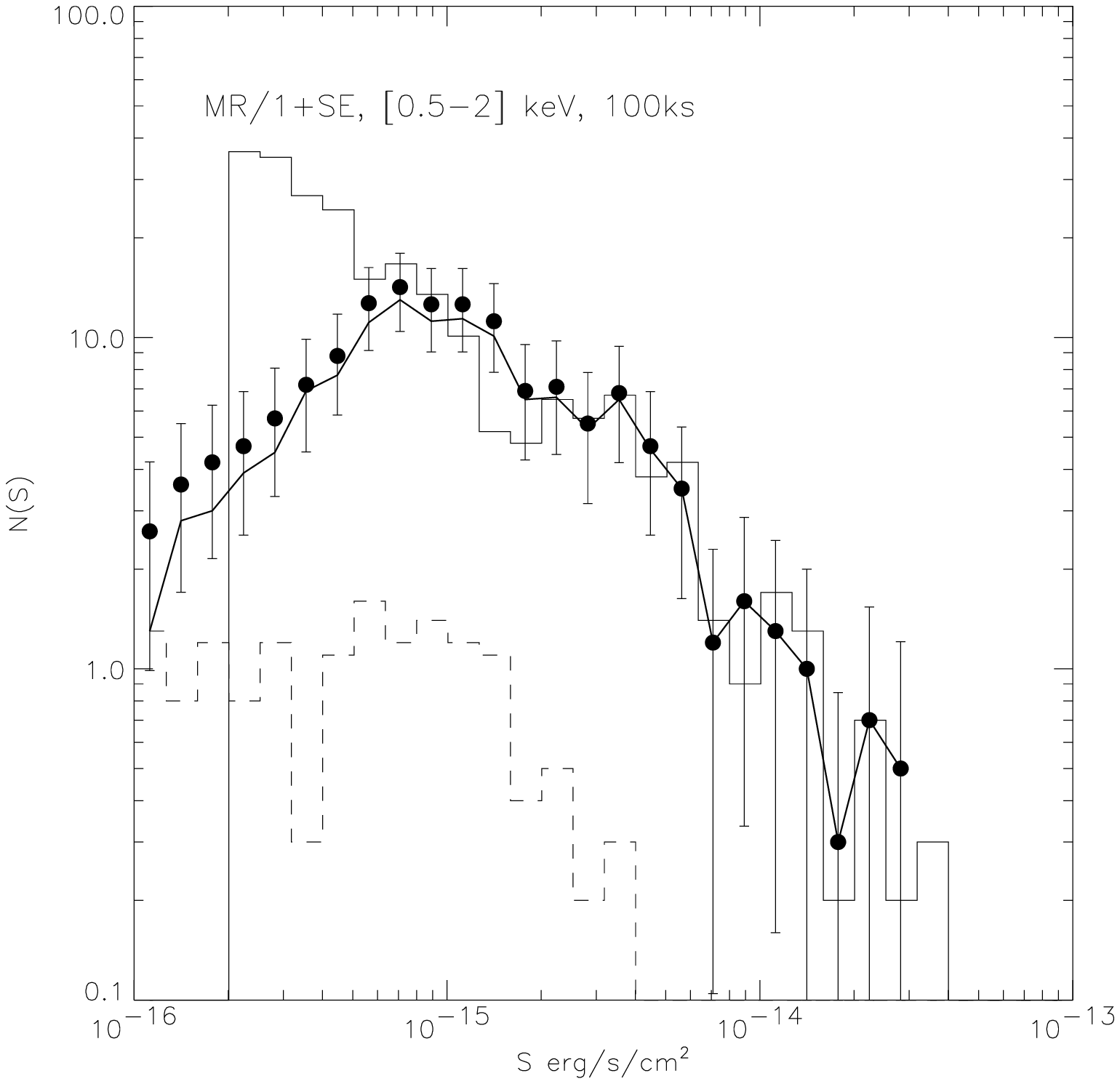} \hfill
        \includegraphics[width=7cm]{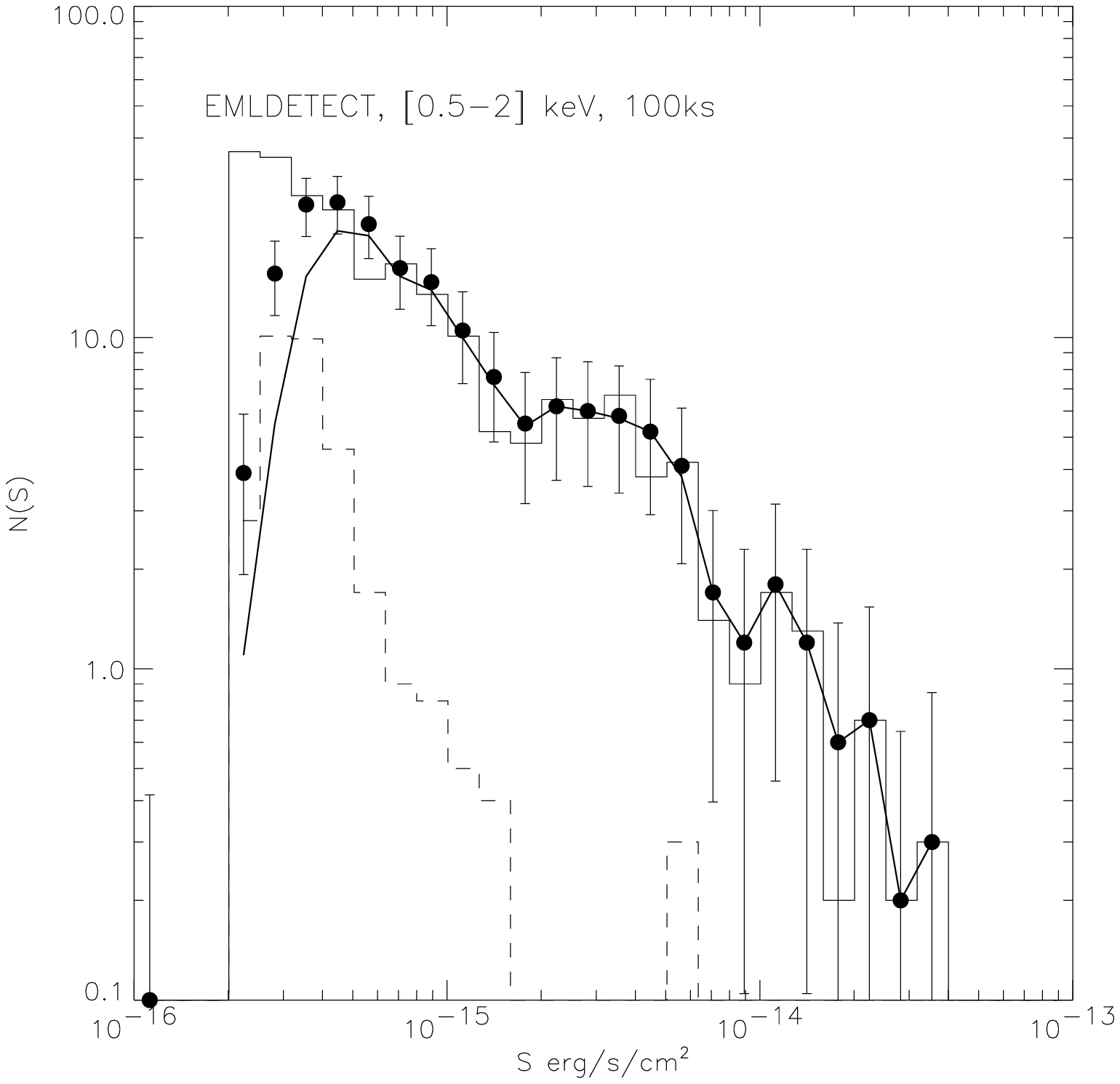} }}
    \caption{
      Differential number counts as a function of flux. The continuous
      histogram is the input distribution, the dots with error bars
      (Poissonian) are the measured distribution (without any
      cross-identifications) and the dashed histogram indicates the
      false detections -- all are averaged from 10 simulations.  Left
      for \mrsex, right for \eml. }
    \label{test5:diff}
  \end{figure*}
  
\item $\log N-\log S$ \\
  The $\log N-\log S$ functions in [0.5-2] keV for 10 ks and 100 ks
  exposures are shown in Fig.~\ref{test5:integ}.  The inferred $\log N
  - \log S$ (by simple source counting) are in very good agreement
  with the input ones down to fluxes about $2\times 10^{-15}$ and
  $6\times 10^{-16}$ \flux\ with \mrsex and $10^{-15}$ and $3\times
  10^{-16}$ \flux\ for \eml for 10 ks and 100 ks respectively.
  Although there are confusion and completeness problems for 100 ks
  starting at $5-6\times 10^{-16}$ \flux as discussed above, their
  effect is completely masked in the \eml\ integral distribution,
  which seems to be in very good agreement down to $3\times 10^{-16}$
  \flux (the lower flux limit for the Lockman Hole Deep Survey
  analysis of Hasinger et al. \cite{has01}).

  \begin{figure*}
    \centerline{\hbox{
        \includegraphics[width=7cm]{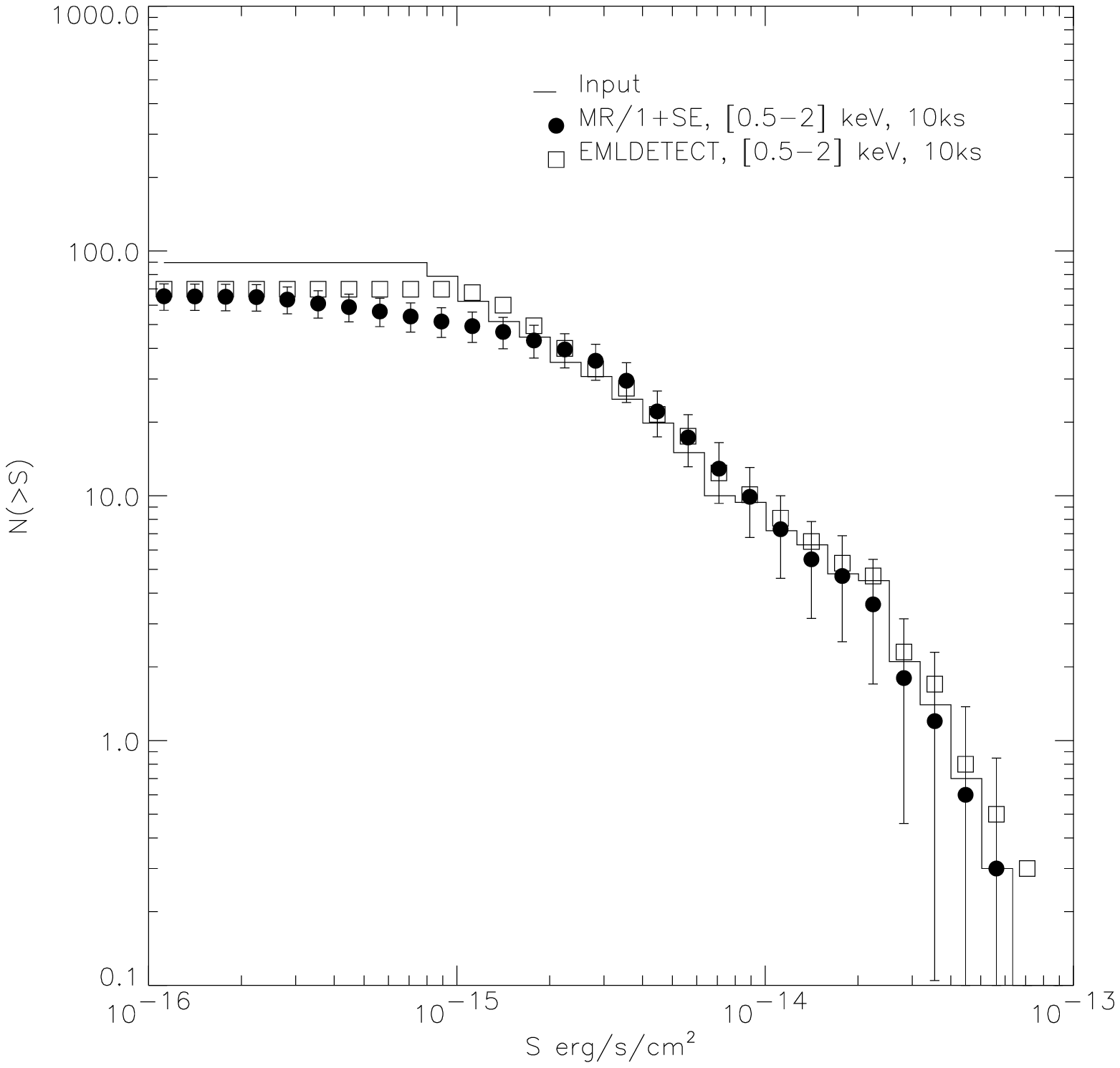} \hfill
        \includegraphics[width=7cm]{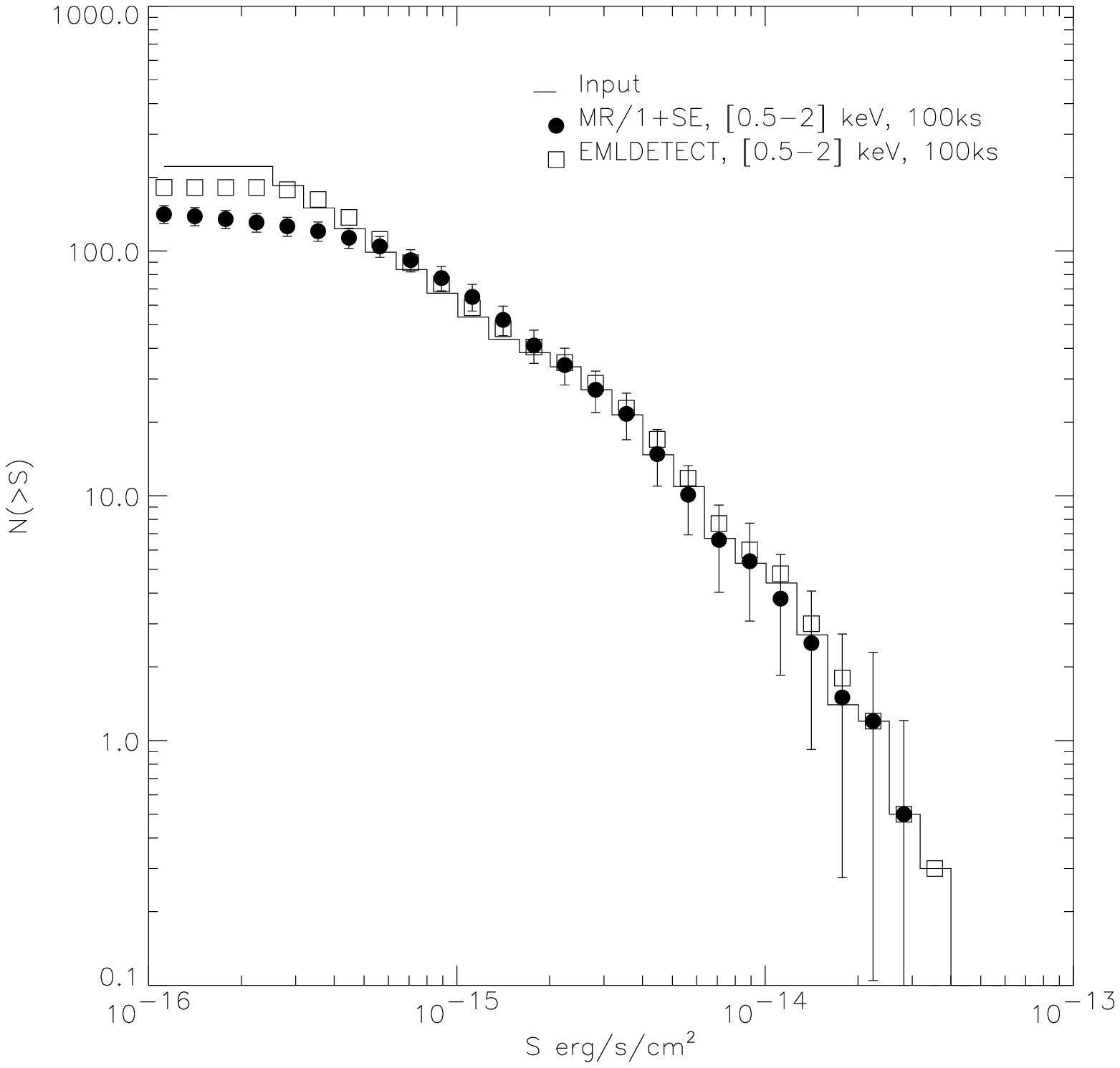}
        }}
    \caption{
      The integral number of objects in the inner $10\arcmin$ (surface
      0.087 sq.deg.) as a function of the flux for [0.5-2] keV and two
      exposures: 10 ks and 100 ks. The points with error bars
      (Poissonian) are the detections (without cross-identification)
      with \mrsex\, boxes are \eml results, while the histogram is the
      input $\log N-\log S$ function.}
    \label{test5:integ}
  \end{figure*}

\item {\it Photometric accuracy}\\
  The photometric reconstruction is relatively similar for 10ks and
  100ks in for fluxes greater than $2\times 10^{-15}$ and $6\times
  10^{-16}$ \flux\ respectively (the 90\% completeness limit,
  Fig.~\ref{test5:photo}). The flux uncertainties are about 25-30 \%
  and going down to fainter fluxes leads to very poor photometry.
  
  \begin{figure*}
    \centerline{\hbox{
        \includegraphics[width=7cm]{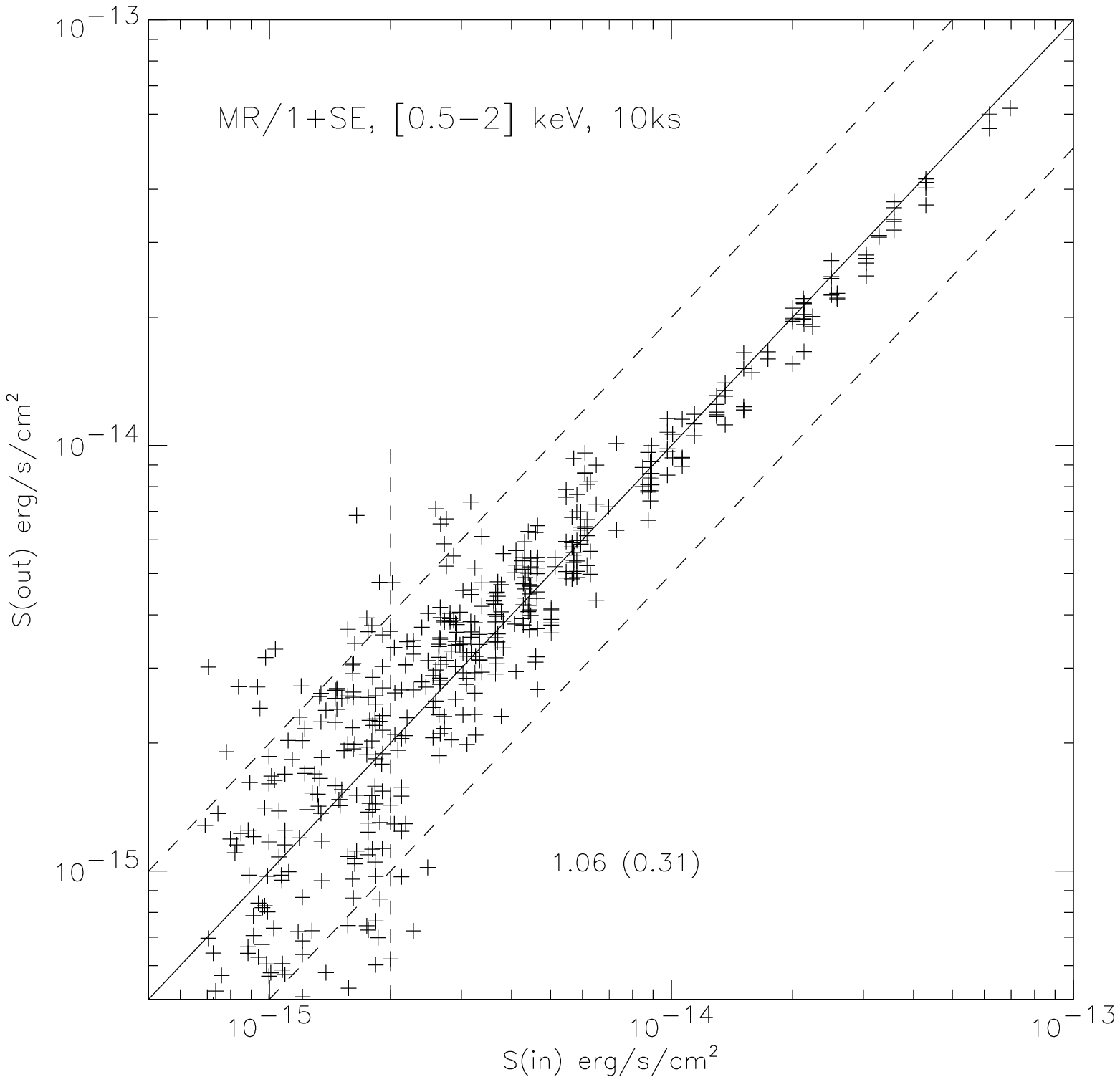} \hfill
        \includegraphics[width=7cm]{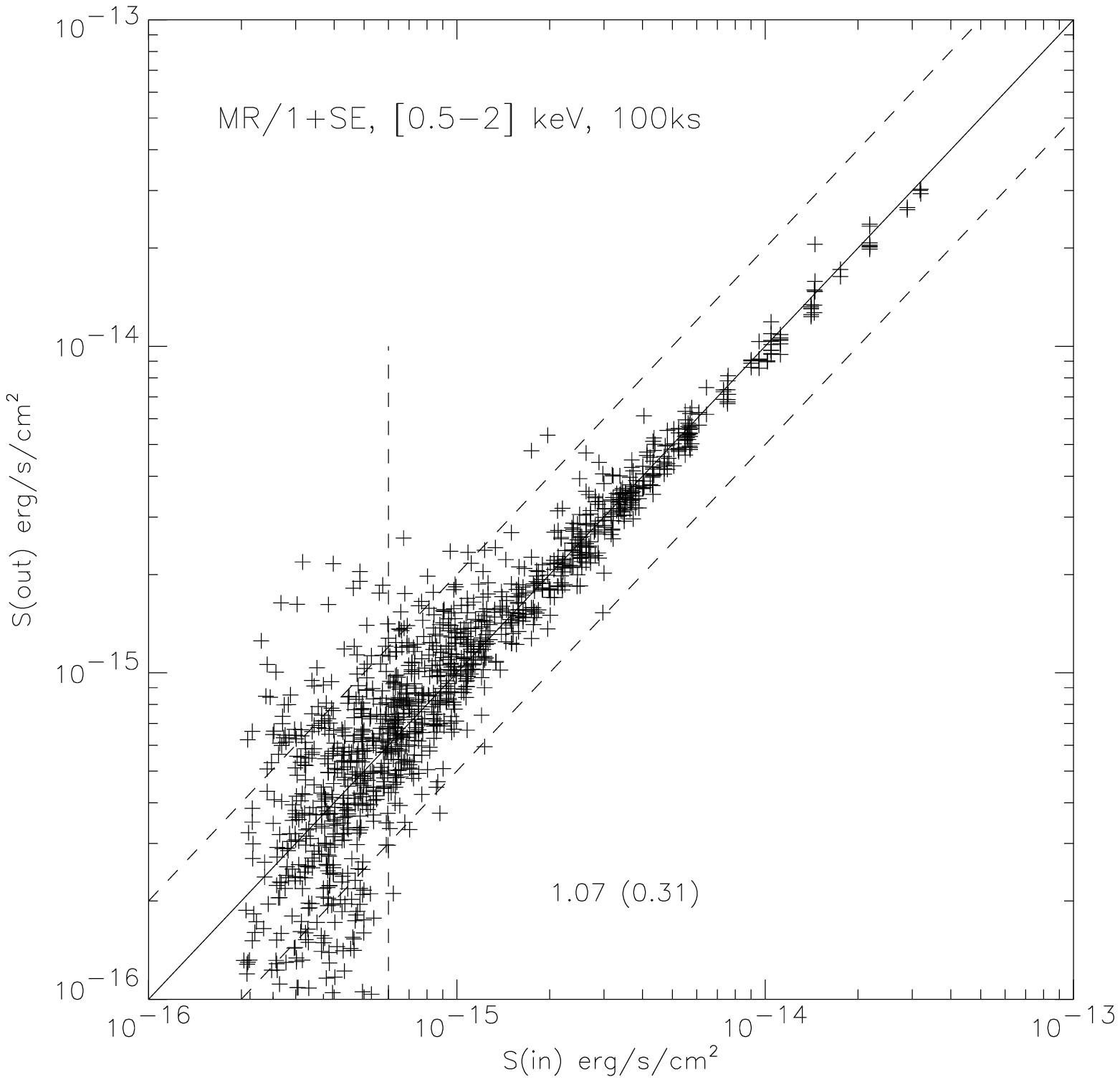}
        }}
    \caption{
      Photometry reconstruction for all 10 simulated images at 10 ks
      (left) and 100 ks (right) in the [0.5-2] keV band. The solid
      line is exact match between detected and input counts while the
      dashed lines are for two-fold differences. The vertical dashed
      line marks the 90\% completeness limit (see
      Tab.~\ref{tab:test5}) and mean and st.dev. (in brackets) above
      this limit are denoted.}
    \label{test5:photo}
  \end{figure*}

\end{itemize}

\section{Conclusions}
\label{sec:disc}

Various procedures for detecting and characterizing sources were
tested by simulated X-ray images. We have concentrated our attention
mainly on images with \XMM specific characteristics, because the
problems arising from its high sensitivity and relatively large PSF
are new and challenging.

We have analyzed the detection rate and the recovery of all
characteristics of the input objects: flux, positional accuracy,
extent measurements and the recovery of the input $\log N-\log S$
relation. We have also investigated confusion problems in large
exposures.

Concerning detection rate and characteristics reconstruction, we have
shown that the \vtp\ implementation of the Voronoi Tessellation and
Percolation method is not suited to \XMM images. \ewav\ provides very
good detection rate and photometric reconstruction for point-like
sources after a simple correction, but shows unreliable results for
extended sources.

One of the best methods for point-like source detection and flux
measurements is \eml\ but we stress again that the PSF model used for
the ML procedure needs to be close to the image PSF for the most
accurate photometry.  Serious drawbacks are the relatively large
number of spurious detections as well as the splitting of the extended
sources, which we were not able to suppress even with 6 simultaneous
PSF profile fits in the multi-PSF mode; this seriously hampers the
analysis of the extended sources.

\wav\ is a flexible method giving good detections even in some
complicated cases. But, here again, spurious detections are quite
numerous.  \wav\ does not assume a PSF model but requires the PSF size
as a function of the encircled energy fraction and the off-axis
distance in order to define the object detection cell. However, the
way the detection cell is defined leads to bad photometry for
extended objects.

Our choice is the \mrsex\ method. The mixed approach involving first a
multiresolution iterative threshold filtering of the raw image
followed by detection and analysis with \sex. Our tests have shown
that this is the best strategy for detecting and characterizing both
point-like and extended objects . Even though this mixed approach
consists of two distinct steps, it is one of the fastest procedures
(Tab.~\ref{tab:cpu}), allowing easy checks of different stages in the
analysis (filtering, detection, photometry).

\begin{table}  
\caption[]{
  CPU times for performing Test~5 for 100ks. The
  procedures were run on a Pentium III-Xeon, 550MHz, Linux. \gsex\  and
  \mrsex\  CPU time is the total of filtering and detection passes. For
  \eml\  the number of input objects is given.} 
  \label{tab:cpu}
  \begin{tabular} {ccc}
    \hline
    Procedure & Number of & CPU time \\
              &  detections & [min.] \\
    \hline\hline
    \eml      & 528   & 12.0      \\
    \ewav     & 364   & 0.4       \\
    \mrsex      & 370   & 1.9  \\
    \gsex       & 365   & 0.1  \\
    \wav        & 378   & 10.3       \\
    \vtp        & 1307 & 10.7        \\
    \hline
  \end{tabular}
\end{table}

Without blending or confusion effects, the photometry is accurate
within 10-20\% for both point-like and extended objects.  This
uncertainty can be regarded as an intrinsic error due to the
Poissonian nature of the X-ray images. For extended objects, only the
\mrsex\ method gives satisfactory photometric results.

Blending between extended and point-like sources is quite serious at
separations below $\sim 30\arcsec$.  Better results for photometry may
eventually be obtained if the intrinsic shape of the extended objects
is known, and if the two objects are detected. However, in most of the
cases with small separation, there is no indication of blending --
which is a dangerous situation for flux reconstruction.  In such
cases, there may exist some spectral signatures of the effect.

The identification process of X-ray sources relies on their positional
accuracy.  We have shown that for objects with more than 100 counts in
10 ks exposure images and within the inner $10\arcmin$ of the
field-of-view, the one sigma positional error is of the order of one
half of the FWHM of the PSF ($\approx 3 - 4\arcsec$,
Tab.~\ref{tab:dr}).  For extended objects, because of their shallower
profiles and depending on the number of photons and the off-axis
distance, the detected centre could even be at about $15\arcsec$ from
its input position.

Comparing series of simulations with 100 ks and 10 ks in two energy
bands -- $[0.5-2]$ and $[2-10]$ keV, we show that the effects of
confusion and completeness are absent for 10 ks, but quite significant
for 100 ks in the lower energy band. Moreover, for faint fluxes, these
effects tend to be masked by the large number of spurious detections
with \eml. Although this method seems to give correct results for the
$\log N- \log S$ down to fainter fluxes than \mrsex, in real
situations it is impossible to asses the contribution of the numerous
spurious detections. From our simulations, we estimate that about
60-65\% of the sources are lost between $3\times 10^{-16}$ and $6
\times 10^{-16}$ \flux\ for a 100 ks exposure with the current best
method (\mrsex).

One of the most important conclusions that will have deep cosmological
impact concerns the detection and classification of extended objects.
We have shown that the \mrsex\ mixed approach is capable of detecting
galaxy cluster-like objects with moderate luminosity ($L_{[2-10]keV}
\sim 3\times 10^{44}$ erg/s) at redshifts $1.5 < z< 2$ in 10 ks \XMM
simulated images. A criteria based on the half-light radius and the
stellarity index classifies them correctly, with a confidence level
greater than 98\%.

\begin{acknowledgements}
  We are thankful to J.-L. Starck for many discussions regarding
  wavelet filtering and detections and for the {\tt MR/1} software, R.
  Ogley and A. Refregier for valuable comments on the manuscript, H.
  Bruner and J.  Ballet for comments and help on XMM-SAS and \eml, E.
  Bertin for help on \sex. We thank also the referee for valuable
  comments and suggestions on the manuscript.
\end{acknowledgements}


\begin{thebibliography}{}
\bibitem[2000]{next}
  Andreon S., Gardiulo G., Longo G., et al, 2000, MNRAS, 319, 700
  (NExt) 
\bibitem[1996]{xspec}
  Arnaud K.A., 1996,  in ASP Conf. Ser., Vol. 101, Astronomical Data
  Analysis Software and Systems V,eds. Jacoby G.H. \& Barnes J. (San
  Francisco: ASP), 17 (XSPEC)
\bibitem[2000]{asch00}
  Aschenbach B., Briel U., Haberl F., et al., 2000, SPIE 4012, 731
  (astro-ph/0007256)  
\bibitem[1996]{sex}
  Bertin E., Arnouts S., 1996, A\&AS 117, 393 (\sex)
\bibitem[1996]{b96}
  Brunner H., 1996, Technical Note (SSC-AIP-TN-0001),
  \mbox{http://xmmssc-www.star.le.ac.uk/documents/}
\bibitem[1976]{cff76}
  Cavaliere A., Fusco-Femiano R., 1976, A\&A 49, 137
\bibitem[1988]{ml88}
  Cruddace R.G., Hasinger G.R., Schmitt J.H., 1988, In: Astronomy from
  large databases, eds. F. Murtagh \& A. Heck (Garching:ESO)
\bibitem[1991]{cru91}
  Cruddace R.G., Hasinger G.R., Tr\"umper J. et al., 1991,
  Exp.Astron. 1, 365
\bibitem[1997]{dam97}
  Damiani F., Maggio A., Micela G., Sciortino S., 1997, ApJ 483, 350
\bibitem[1997]{sdg97}
   De Grandi S., Molendi S., B\"horinger H., et al., 1997, ApJ 486, 738
\bibitem[1999]{ciao}
  Dobrzycki A., Ebeling H., Glotfelty K. et al., 1999, In \Chandra
  DETECT User Guide: http://asc.harvard.edu/
\bibitem[2000]{calbook}
  Erd C., Gondoin P., Lumb D., et al., 2000, XMM Calibration Access
  and Data Handbook, XMM-PS-GM-20 Issue 1.1,
  http://xmm.vilspa.esa.es/calibration/ 
\bibitem[2000]{gia01}
  Giacconi R., Rosati P., Tozzi P. et al., 2000, astro-ph/0007240v2
\bibitem[1990]{emss}
  Gioia I.M., Maccacaro T., Schild R.E., et al., 1990, ApJSS 72, 567
\bibitem[1995]{gre95}
  Grebenev S.A., Forman W., Jones C., Murray S., 1995, ApJ 445, 607
\bibitem[1993]{vtp1}
  Ebeling H., 1993, Ph.D. thesis, MPE Report 250.
\bibitem[1993]{vtp2}
  Ebeling H., Wiedenmann G., 1993, Phys. Rev. 47, 704
\bibitem[2000]{warps}
  Ebeling H., Jones L.R., Perlman E., et al., 2000, ApJ 534, 133 
\bibitem[1996]{fre96}
  Freeman P., Kashyap V., Rosner R. et al., 1996,  in ASP Conf. Ser.,
  Vol. 101, Astronomical Data Analysis Software and Systems V,
  eds. Jacoby G.H. \& Barnes J. (San Francisco: ASP), 163
\bibitem[1993]{has93}
  Hasinger G., Burg R., Giacconi R., et al., 1993, A\&A, 275, 1 (erratum
   1994, A\&A 291, 348)
\bibitem[1998]{has98}
  Hasinger G., Burg R., Giacconi R., et al., 1998, A\&A, 329, 482
\bibitem[2001]{has01}
  Hasinger G., Altieri B., Arnaud M., et al. 2001, A\&A, 365, L45
\bibitem[1987]{inf87}
  Infante L., 1987, A\&A 183, 177
\bibitem[2000]{kol00}
  Kolaczyk E., Dixon D., 2000, ApJ 534, 490
\bibitem[1980]{kro80}
  Kron R.G., 1980, ApJS 43, 305
\bibitem[1988]{inventory}
  Kruszewski, A., 1988, in 1st. ESO/ST-ECF Data Analysis Workshop,
  eds. Grosbol P.J., Murtagh F., Warmels R.H., ESO Munchen, 29
\bibitem[1999]{l99}
  Lazzati D., Campana S., Rosati P., et al., 1999, ApJ 524, 414
\bibitem[2000]{mp00}
  Pierre M., Bryan G., Gastaud R., 2000, A\&A 356, 403
\bibitem[1977]{rs77}
  Raymond J.C., Smith B.W., 1977, ApJS 35, 419
\bibitem[1997]{s97}
  Scharf C.A., Jones L.R., Ebeling H., et al., 1997, ApJ 477, 79
\bibitem[1993]{sle93}
  Slezak E., de Laparent V., Bijaoui A., 1993, ApJ 409, 517
\bibitem[1994]{sle94}
  Slezak E., Durret F., Gerbal D., 1994, AJ 108, 1996
\bibitem[1998]{sp98}
  Starck J.-L., Pierre M., 1998, A\&AS 128, 397
\bibitem[1998]{sta98}
  Starck J.-L., Murtagh F., Bijaoui A., 1998, Image Processing and Data
  Analysis: The Multiscale Approach, Cambridge Univ. Press, Cambridge
  (UK) (\mr) 
\bibitem[1990]{focas}
  Valdes F.G., Campusano  L.E., Velasquez J.D., Stetson P.B., PASP
  107, 1119 (FOCAS)
\bibitem[1997]{vik97}
  Vikhlinin A., Forman W., Jones C., 1997, ApJ 474, L7 
\bibitem[1999]{rass}
  Voges W., Aschenbach B., Boller Th., et al., 1999, A\&A 349, 389
\bibitem[2001]{wat00}
  Watson M.G., Augu\`eres J.-L., Ballet J., et al. 2001, A\&A, 365,
  L51 
\bibitem[1994]{exsas}
  Zimmermann H.U., Becker W., Belloni T., et al., 1994, EXSAS User's
  Guide, MPE Report 257  (EXSAS)
\end{thebibliography}
\end{document}